\documentclass{article}

\usepackage{algorithm}
\usepackage{algpseudocode}
\usepackage{PRIMEarxiv}
\usepackage{algorithm}
\usepackage{algpseudocode}
\usepackage[utf8]{inputenc} 
\usepackage[T1]{fontenc}    
\usepackage{hyperref}       
\usepackage{url}            
\usepackage{booktabs}       
\usepackage{amsfonts}       
\usepackage{nicefrac}       
\usepackage{microtype}      
\usepackage{lipsum}
\usepackage{amsmath, amssymb, amsthm}
\usepackage{hyperref}
\usepackage{fancyhdr}       
\usepackage{graphicx}       
\graphicspath{{media/}}     

\title{Application of Kalman Filter in  Stochastic Differential Equations

}
\author{
  Wencheng Bao, Shi Feng \\
  Department of Applied Physics and Applied Mathematics, Graduate candidate \\
  Columbia University, New York, NY \\
  \texttt{wbao5@illinois.edu}, \texttt{sf3164@columbia.edu}
  \And
  Kaiwen Zhang \\
  Department of Applied Physics and Applied Mathematics, Undergraduate candidate\\
  Columbia University, New York, NY \\
  \texttt{ kz2387@columbia.edu} 
}

\begin{document}
\maketitle
\begin{abstract}
In areas such as finance, engineering, and science, we often face situations that change quickly and unpredictably. These situations are tough to handle and require special tools and methods capable of understanding and predicting what might happen next. Stochastic Differential Equations (SDEs) are renowned for modeling and analyzing real-world dynamical systems. However, obtaining the parameters, boundary conditions, and closed-form solutions of SDEs can often be challenging. In this paper, we will discuss the application of Kalman filtering theory to SDEs, including Extended Kalman filtering and Particle Extended Kalman filtering. We will explore how to fit existing SDE systems through filtering and track the original SDEs by fitting the obtained closed-form solutions. This approach aims to gather more information about these SDEs, which could be used in various ways, such as incorporating them into parameters of data-based SDE models.
\end{abstract}

\keywords{Kalman Filter \and Extended Kalman Filter \and Particle Kalman Filter \and Heston \and Stochastic Differential Equations }

\section{Introduction}

Stochastic Differential Equations (SDEs) are formed by adding one white noise term to an ordinary differential equation. These equations are particularly useful in modeling systems or processes that exhibit random behaviors. In real life, SDEs have many applications, such as the Black-Scholes Model\cite{LAUTERBACH1990} used in option pricing, the Fokker-Planck Equation\cite{doi:10.1137/S0036141096303359} in physics describing the time evolution of probability densities under random forces, and the Lotka-Volterra Equations with Noise\cite{DU200682} in describing classic predator-prey relations.

However, not all SDEs have closed-form solutions. Additionally, there arises the question of how to apply known SDEs to actual data, including how to select the most appropriate parameters and set the appropriate boundary conditions. These questions have been studied up to the present day. For example, the Heston model is known as a model for option pricing \cite{heston}. To solve parameter estimation in the Heston model, various methods have been tried, including Maximum Likelihood Estimation (MLE) based on historical equity prices, the Efficient Method of Moments (EMM) algorithm, the Monte Carlo Markov Chain (MCMC) algorithm, and Quasi-Maximum Likelihood Estimation (QMLE)\cite{WANG201714100}. Nevertheless, when it comes to capturing elusive factors like unknown volatility, filtering theory has emerged as a pivotal approach, offering new avenues for understanding and managing the complexities inherent in SDEs.

The filtering theory studies the complex challenges of estimating systems, as the estimation system itself is incomplete and susceptible to noise interference\cite{1055174}. This field is mainly applied in two key situations: first, direct observation of variables is impractical and requires indirect measurement; second, it is necessary to integrate multiple variables, each of which is prone to random fluctuations. This article aims to focus on the Kalman filtering and its various extensions, exploring their roles and significance in the above SDEs. By applying these advanced filtering methods, we can approximate the behavior of SDEs even in the absence of neat, closed-form solutions, offering an approach to understanding and navigating the stochastic dynamics of real-world phenomena.

\section{Technique Simulation Review}
On the problem of learning and fitting time-series data with uncertainties, there is abundant literature using advanced computer tools and programs to perform the task. These tools include various types of deep neural networks, including multi-layer perceptron (MLP)\cite{app12031357}, recurrent neural networks such as gated recurrent unit (GRU)\cite{ARUNKUMAR20227585} and long short term memory (LSTM)\cite{Song2020}), and transformers\cite{Kotb2022}, which are in general well-suited for identifying complex patterns. There are also papers using reinforcement learning methods such as parameter-SARSA\cite{ZHU2022107658} and parameter-TD\cite{Bradtke1996} in parameter learning problems. However, we found that these methods might not be as effective in the context of SDEs. Consider the results presented in  $\bf{Figure}$ $\bf{\ref{fig:1}}$, for example, where we simulated sequential data following the Heston Model $\bf{Equation}$ $\bf{\ref{eq:heston}}$ developed in \cite{heston}, and then attempted to fit it with neural networks:

\begin{figure}[h!]
\begin{center}
\includegraphics[scale=.16]{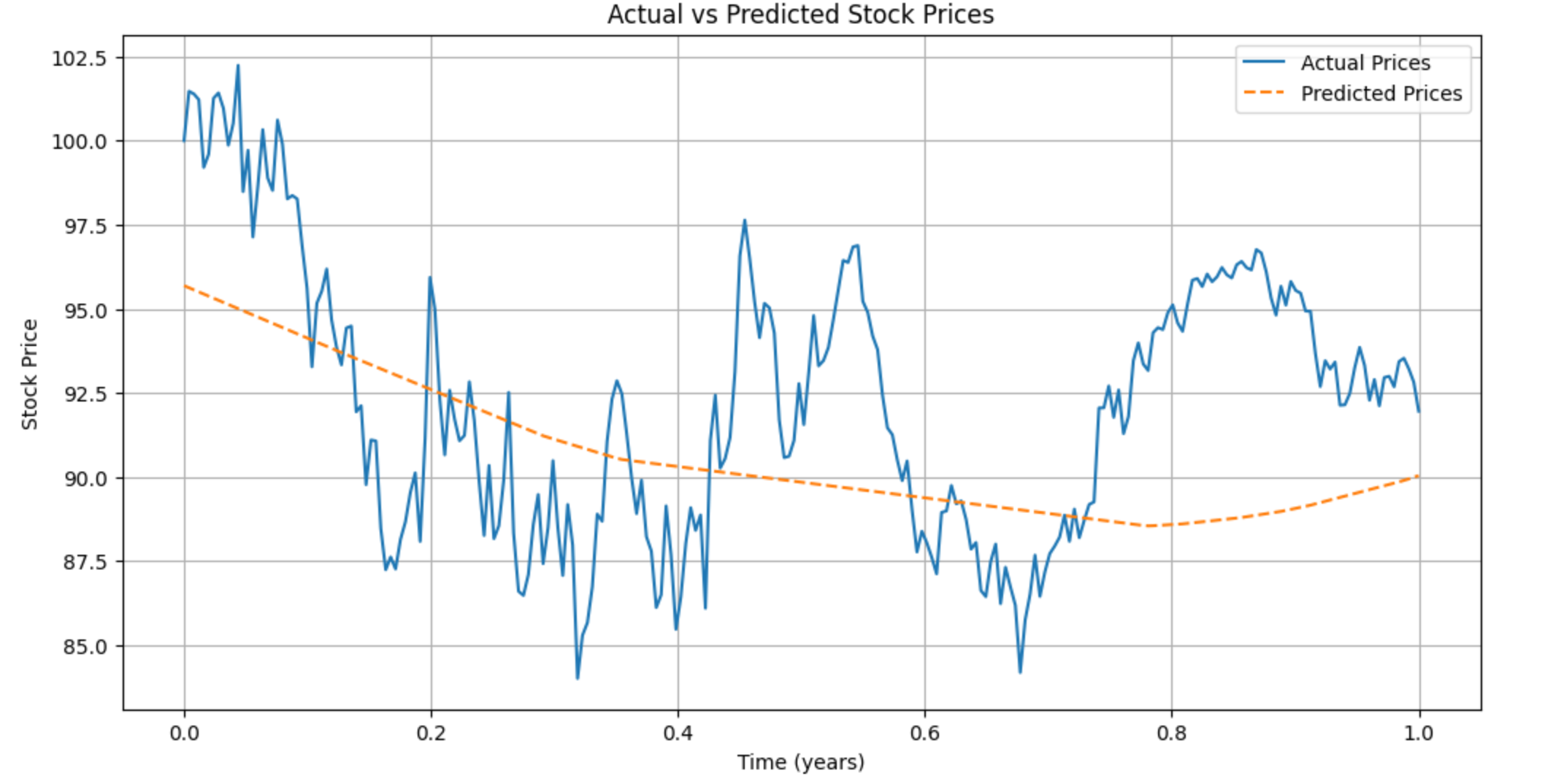}
\includegraphics[scale=.16]{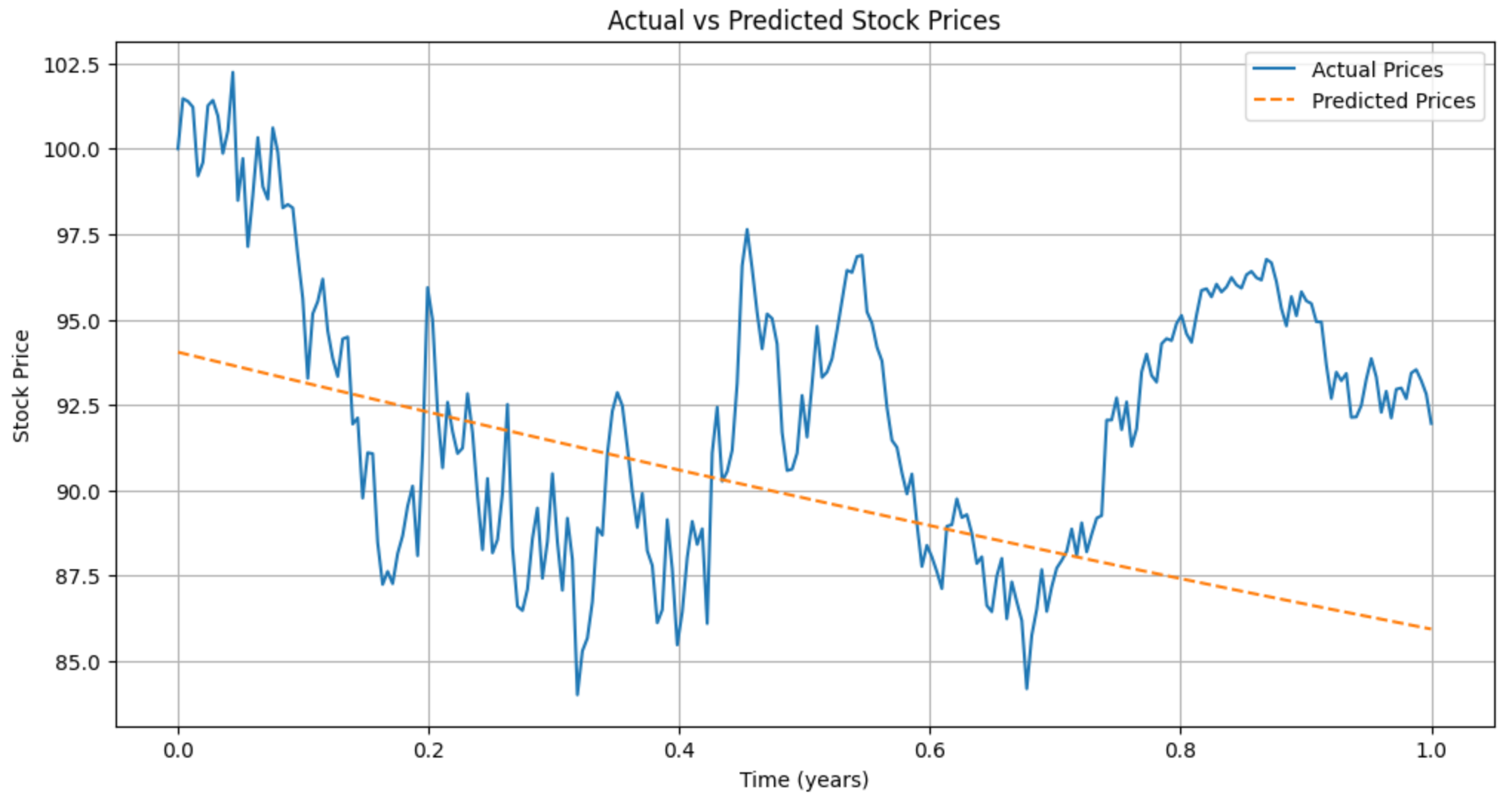}
\caption{\label{fig:1}Neural Network to Approach a Heston Model} 
\end{center}
\end{figure}

In the upper figure we used MLP. The model is a sequential neural network with six layers. It begins with a dense layer of 128 neurons (ReLU activation, input size 1), followed by a 20\% dropout. The next layer repeats this setup with another 128-neuron dense layer (ReLU) and a 20\% dropout. It then narrows down to a 64-neuron dense layer (ReLU), a 32-neuron dense layer (ReLU), and concludes with a single-neuron dense output layer. This structure is designed for complex pattern recognition and regression tasks. The dropout layers were included to reduce overfitting. 

In the lower figure, we used another neural network, also designed for sequential data, but starting with an input layer shaped for time series. It features an LSTM layer with 128 units for capturing temporal dependencies, enhanced by an attention mechanism focusing on crucial sequence parts. The output from attention mechanism is flattened, leading to two dense layers (64 and 32 neurons, both with ReLU activation), separated by a 20\% dropout layer to reduce overfitting. The network culminates in a single-neuron output layer, making it suitable for tasks requiring attention to specific time-dependent patterns.

We trained the models for approximately 50 epochs, and found that the two neural networks cannot fit the data well and are computationally very expensive. Therefore, we sought for more efficient and effective mathematical approaches to the problem of learning information within the realm of stochastic differential equations. In this paper, we will introduce Kalman filtering, a signal processing method constructed through state transition equations, and its application to parameter estimation in various SDEs. Kalman Filter will provide closed-form solutions for the estimation of states.

\section{Maximum Likelihood Estimation (MLE) Framework}

Parameter estimation is framed as an inverse problem, involving the deduction of system-characterizing parameters from observational data, which is presumed to follow the system's rules. In scenarios where we encounter noisy data, it is assumed that this data aligns with an equation of a predetermined form, but the coefficients remain unknown.

\subsection{Likelihood Function}

Maximum Likelihood Estimation (MLE) is a technique for parameter estimation when the target to estimate involves a probability distribution. In general, let $D$ be a probability distribution with parameter $\theta$, to be estimated. Given observations $x$ sampled from $D$, one can compute the probability of seeing $x$ as a function of $\theta$, $P(x \vert \theta)$. Maximizing this probability yields an optimal choice of $\theta$, called maximum likelihood estimate. One could denote the function $\mathcal{P}(\theta) = P(x \vert \theta)$ as the "likelihood" function of $\theta$, although in continuous settings one need not maximize this function explicitly in this form.

In the context of MLE applied to stochastic differential equations (SDEs), the data given is a discrete time-series observation: $X: \{0, 1, ..., N\} \to \mathbb{R}$ or $X = (X_0, X_1, ..., X_N)$. Suppose the SDE contains an unknown parameter $\theta$. Each step in the observation depends on the preceding step. Hence, the probability of seeing an observation is: 
\begin{equation}
P(X \vert \theta) = P(X_0 \vert \theta) \cdot P(X_1 \vert X_0;\theta) \cdot P(X_2 \vert X_1;\theta) ... P(X_N \vert X_{N-1}; \theta)
\end{equation}

If $X_t$ for each $t$ follows a continuous conditional distribution, then $P(X_t \vert X_{t-1};\theta) = f(X_t \vert X_{t-1}; \theta) dx$ where $f$ is the probability density function. Notice then that it suffices to drop the $dx$ and compare the product of probability density function values, instead of the probabilities themselves. This is in turn our "likelihood function", defined as:
\begin{equation} \label{eq:likelihoodFunc}
    L(\theta) = f(X_0) f(X_1 \vert X_0; \theta) \cdot f(X_2 \vert X_1; \theta) ... f(X_N \vert X_{N-1}; \theta) = f(X_0) \prod_{t = 1}^{N} f(X_t \vert X_{t-1}; \theta)
\end{equation}
The goal in an MLE, then, is to compute and maximize $L(\theta)$. In practice, the precise formula could overflow, so instead we maximize the log-likelihood function. Simply take the logarithm of $\bf{Equation}$ $\bf{\ref{eq:likelihoodFunc}}$ to define:
\begin{equation} \label{eq:LogLikelihood}
    \ell(\theta) = \log(f(X_0)) + \sum_{t=1}^{N}\log(f(X_t \vert X_{t-1}; \theta))
\end{equation}

\subsection{Conditional Density for Itô Diffusion}

Suppose observation $ {X} = (X_0, X_1, ..., X_N)$ follows an Itô diffusion of form and
\begin{equation} \label{eq:ItôDiffCont}
    dX_t = \mu(X_t, t, \theta) dt + \sigma(X_t, t, \theta) dW_t
\end{equation}

Discretizing $\bf{Equation}$ $\bf{\ref{eq:ItôDiffCont}}$ gives rise to the following analogue:
\begin{equation} \label{eq:ItôDiffDisc}
    X_{t} - X_{t-1} = \mu(X_{t-1}, t-1, \theta) d_{t-1} + \sigma(X_{t-1}, t-1, \theta) (W_{t} - W_{t-1})
\end{equation}

where $dt$ is the time step. We study the conditional density $f(X_t \vert X_{t-1}; \theta)$. By definition of Brownian motion, given the observation $X_{t-1}$, $X_{t}$ follows a normal distribution centered at $X_{t-1} + \mu(X_{t-1}, t-1, \theta) d_{t}$, with variance $(\sigma(X_{t-1}, t-1, \theta))^2 \cdot dt$. Hence we can write down the formula:

\begin{align} \label{eq:ItoCondDensity}
    f(X_t \vert X_{t-1}; \theta) = \frac{1}{\sqrt{2\pi dt}\sigma(X_{t-1}, t-1, \theta)} \exp\left\{-\frac{1}{2}\frac{(X_t - X_{t-1} - \mu(X_{t-1}, t-1, \theta) dt)^2}{(\sigma(X_{t-1}, t-1, \theta))^2 \cdot d_{t}}\right\}
\end{align}

To maximize $\bf{Equation}$ $\bf{\ref{eq:likelihoodFunc}}$ with the conditional density given by $\bf{Equation}$ $\bf{\ref{eq:ItoCondDensity}}$ or $\bf{Equation}$ $\bf{\ref{eq:LogLikelihood}}$. 








\subsection{Marginal likelihood}
The Kalman filter, akin to the Hidden Markov Model, differs in its use of Gaussian distributed continuous variables instead of discrete states and observations \cite{18626}. It's important to assess the probability of the filter producing a specific observed signal based on its parameters, such as prior distribution, transformation, observation model, and control inputs. This probability, known as the marginal likelihood, integrates over hidden state variables, allowing computation with just the observed signal. It's valuable for evaluating parameter setups and comparing the Kalman filter with other models using Bayesian methods.

The Kalman filter encapsulates a Markov process, wherein all pertinent information from prior observations is encapsulated within the current state estimate. The Marginal likelihood can be taken as the product of the probabilities of each observation given the previous observations according to the chain rule. If we have random observation state $(z_1, z_2,..., z_t,...)$, following the regular Kalman filter System ($H_t$ is the observation model, $R_t$ is the covariance of the observation noise, and $P_t$ is the a posteriori estimate covariance matrix), then:
\begin{equation} 
f(z) =\prod_{t=0}^N \int f\left(z_t \mid X_t\right) p\left(X_t \mid z_{t-1}, \ldots, z_0\right) d X_t =\prod_{t=0}^N \mathcal{N}\left(z_t ; H_t \hat{X}_{t \mid t-1}, R_t+H_t P_{t \mid t-1} H_t^{\top}\right) 
\end{equation}

Consider $t=0$, we have a series of Gaussian densities, each linked to an observation $z_t$ under the filtering distribution $H_t \hat{x}_{t \mid t-1}, R_t+H_t P_{t \mid t-1} H_t^{\top}$. A recursive update simplifies this process. To avoid numerical underflow, the log marginal likelihood, $\ell=\log p(z)$, with $\ell^{(-1)}=0$ , is computed instead. This is achieved through a recursive formula, considering $d_y$ represents the dimension of the measurement vector \cite{MarginallikelihoodFunction}:
\begin{equation}\label{eq:marginallikeihood}
\ell^{(t)}=\ell^{(t-1)}-\frac{1}{2}\left(\tilde{y}_t^{\top} (R_t+H_t P_{t \mid t-1} H_t^{\top})^{-1} \tilde{y}_t+\log \left|R_t+H_t P_{t \mid t-1} H_t^{\top}\right|+d_y \log 2 \pi\right)
\end{equation}

\section{Kalman Filter Framework}
One topic in control theory is to accurately estimate the state of a dynamical system. Often, the state produces an observable measurement. The simplest setup of such a system is characterized in discrete updating schemes:
\begin{equation} \label{eq:kf}
\begin{aligned}
    x_{t} &= A_{t-1} x_{t-1} + G_{t-1} W_{t-1} \\
    y_t &= H_t x_t + \epsilon_t
\end{aligned}
\end{equation}

The first equation is the updating scheme for the state variable $x_{i}$, and the second equation represents how the state variable $x_{i}$ produces a measurement $y_{i}$. $W_t$ and $\epsilon_t$ are random variables, representing noises involved in the state update and measurement processes. In our studies, the two noises at the same time step could be correlated, but any noise variables at different time steps are uncorrelated. Both $x$ and $y$ are random variables. One usually refers to their expectations when talking about "estimates".

Kalman filtering is a technique which uses real-time measurements to correct state estimations. Let $Q_t$ be the covariance of process noise $W_t$ and $R_t$ be the covariance of measurement noise $\epsilon_t$. Denote $P_t$ as the a posteriori state estimation error covariance, and $P_t^{-}$ as the a priori state estimation error covariance. \cite{Asada_2006_5} and \cite{Asada_2006_6} describe the algorithm to achieve optimal state estimation. We reproduce it below, 
$K_t$ is called the Kalman gain, which theoretically minimizes the a posteriori estimate.

\begin{algorithm}
\caption{Kalman Filter}
\begin{algorithmic}[1]

\State \textbf{Input:} The system matrices $A_t$, $H_t$, $Q_t$, $R_t$, and the initial state $\hat{x}_0$ and covariance $P_0$.
\State \textbf{Output:} Updated state estimate $\hat{\mathbf{x}}_t$ and error covariance $P_t$.
\For{each time step $t = 1, 2, \ldots$}
    \State Propagate the a priori state estimate:
    $\hat{x}_t^{-} = \mathbb{E}(x_{t|t-1}) = A_{t-1} \hat{x}_{t-1}$

    \State Produce the estimated measurement:
    $\hat{y}_t = \mathbb{E}(y_{t|t-1}) = H_t \hat{x}_t^{-}$

    \State Compute the Kalman gain matrix:
    $K_t = P_t^{-} H_t^{T}[H_t P_t^{-}H_t^{T} + R_t]^{-1}$

    \State Correct the a priori estimate with the actual measurement:
    $\hat{x}_t = \hat{x}_t^{-} + K_t (y_t - \hat{y}_t)$

    \State Update error covariance matrix:
    $P_t = (I - K_t H_t)P_t^{-}$, and
    $P_{t+1}^{-} = A_tP_tA_t^{T} + G_t Q_t G_t^{T}$

\EndFor

\end{algorithmic}
\end{algorithm} 

Delving into the realm of control systems, we can unveil the one-dimensional formulation of the Kalman gain:
\begin{equation}
\begin{aligned}
\hat{x}_t &= \hat{x}_t^{-}+K_t\left(z_t-H_t \hat{x}_t^{-}\right) = \hat{x}_t^{-}+K_t\left(z_t-H_t x_t+H_t e_t^{-}\right)
\\
e_t &= x_t-\hat{x}_t=e_t^{-}-K_t\left(\epsilon_t+H_t e_t^{-}\right)
\end{aligned}
\end{equation}

Then, we update the error covariance matrix:
\begin{equation}
P_t= E\left[e_t^2\right]=P_t^{-}+K_t^2\left(R_t+H_t^2 P_t^{-}+0\right)-2 K_t H_t P_t^{-}
\end{equation}
By differentiating with respect to $K_t$ and ingeniously setting this derivative equal to zero, we arrive at the solution:
\begin{equation}
K_t=\frac{P_t^{-} H_t}{H_t^2 P_t^{-}+R_t}
\end{equation}

\label{sec:headings}
\subsection{Linear Kalman Regression}
In machine learning, linear regression is supervised learning, which assumes that each measurement has the same weight. Kalman filtering incorporates the consideration of internal changes in the system. The process equation is used to predict the system's state at the next moment based on its current state. When we need to fit a known linear system, we can use linear Kalman regression \cite{SORENSON1966219}. By changing the measured noise, system noise will affect the effectiveness of Kalman filtering without affecting the linear regression filtering, while changing the order of least squares will affect it. 

Consider a linear regression defined in the following formula:
\begin{equation}\label{eq:LKRE}
    Y = \alpha + \beta X + \epsilon
\end{equation}
where $Y$ and $X$ are response and predictor variables, respectively. $\epsilon$ is the noise in system, for which $\text{Std}(\epsilon) = \sigma$ and $ \text{Cov}(\epsilon, X) = 0$. In the Kalman Filter setting, the state at time $t$ is: $x_t = \left(\begin{array}{c}x_{\alpha_t} \\ x_{\beta_t}\end{array}\right)$. View the measurement at the next time step as response variable $Y$, and then the system satisfies the following process and measurement equations:
\begin{equation} \label{eq:LinearKalm}
\begin{aligned}
\left(\begin{array}{c}
x_{\alpha_{t}} \\
x_{\beta_{t}}
\end{array}\right) &= A_{t-1}\left(\begin{array}{c}
x_{\alpha_{t-1}} \\
x_{\beta_{t-1}}
\end{array}\right) + G_{t-1}W_{t-1} \text { with } G_{t-1}W_{t-1} \sim \mathcal{N}\left(0, Q_{t-1}\right) \text { and }Q_{t-1}=\left(\begin{array}{cc}
\sigma_\alpha^2 & 0 \\
0 & \sigma_\beta^2
\end{array}\right)
\\
y_t &=H_t\left(\begin{array}{c}
x_{\alpha_t} \\
x_{\beta_t}
\end{array}\right)+\epsilon_t \text { with } H_t = \left(\begin{array}{ll}
1 & x_t
\end{array}\right) \text{ and } \epsilon_t \sim \mathcal{N}\left(0, \sigma_\epsilon^2\right)
\end{aligned}
\end{equation}
Normally, for a simple linear system, we can regard the transit matrix $A_t$ as an identity matrix. And the noise term $G_tW_t = \left(\begin{array}{c}\nu^{\alpha}_t \\ \nu^{\beta}_t\end{array}\right)$, where $\nu^{\alpha}_t$ and $\nu^{\beta}_t$ are independent. Then the rest of the calculation is supposed as the same as the regular Kalman Filter.

In the above $\bf{Equation}$ $\bf{\ref{eq:LinearKalm}}$, the noise term $G_tW_t$ and $\epsilon_t$ contain parameters $Q_t$ and $\sigma_{\epsilon}$. The idea is to calibrate with the real-world data by Maximum Likelihood Estimation (MLE), which will be introduced in Chapter 4. By following the marginal likelihood function $\bf{Equation}$ $\bf{\ref{eq:marginallikeihood}}$:
\begin{equation}\label{eq:logLKR}
\log L\left(\alpha, \beta, Q_k, \sigma_{\epsilon} \mid y_t, y_{t-1}, \ldots, y_1, y_0\right)=-\frac{1}{2} \sum_{i=1}^N\left(\frac{r_t^2}{S_t} + \log S_t + \log 2 \pi\right)
\end{equation}
where $r_t$ is the error between the true and estimated state. $S_t$ is the variance of the system $\bf{Equation}$ $\bf{\ref{eq:LKRE}}$, the system contains mean: $\text{E}[y_{t} | \alpha_{t-1,t-1} 
 \beta_{t-1,t-1}] = \alpha_{t,t-1} + \beta_{t,t-1}X_{t-1} + \epsilon_{t-1}$ and variance: $\text{Var}[y_{t} | \alpha_{t-1,t-1} \beta_{t-1,t-1}] = x_{t}^2 P_{t, t-1}+\sigma_\epsilon^2$.

\subsection{Extended Kalman Filter (EKF)}
In the above discussion, the state and measurement equations are assumed to be linear. However, in reality, these equations are often nonlinear. The Extended Kalman Filter (EKF) is a technique that generalizes Kalman filtering to such settings.

Suppose the state and measurement equations for a dynamical system are:
\begin{equation}
\begin{aligned}
    x_{t} &= f(x_{t-1}, w_{t-1}) \quad w_{t-1} \sim N\left(0, Q_{t-1}\right) \hspace{1em} \text{state}\\
    y_t & = h(y_{t-1}, \epsilon_{t-1}) \quad \epsilon_{t-1} \sim N\left(0, R_{t-1}\right) \hspace{1.3em} \text{measurement}
\end{aligned}
\end{equation}

Where $f, h$ are assumed to be smooth functions. To linearize this system, we formulate a benchmark state trajectory and study the difference between the actual state and the benchmark state. $Q_t$ and $R_t$ is the covariance matrix of two noise terms. In addition, $w_{t-1}$ is uncorrelated with $x_t$ and $\epsilon_{t-1}$ is uncorrelated with $y_t$.

For a function $f(x)$, we know that its first-order Taylor expansion form is:
\begin{equation}
f(x)=f\left(x_k\right)+\left(x-x_k\right) f^{\prime}\left(x_k\right)
\end{equation}
As a state function, x estimates its linear prior process is $\hat{x}_{t}^{-}=E\left[x_t\right]$, and $\hat{x}_{t}=E \left[x_t | y_t\right]$. Assuming that the new input $u_t$ arrives and the true value of the state is $x_{t-1}$, a first-order Taylor expansion is used near the estimated value $\hat{x}_{t-1}$:
\begin{equation}
\begin{aligned}
f\left(x_{t-1}, u_t\right) &\approx f\left(\hat{x}_{t-1}, u_t\right)+f^{\prime}\left(\hat{x}_{t-1}, u_t\right)\left(x_{t-1}-\hat{x}_{t-1}\right)\\
f^{\prime}\left(\hat{x}_{t-1}, u_t\right)&= \partial f(x_{t-1}, u_t) / \partial x_{t-1}
\end{aligned}
\end{equation}
where $f^{\prime}\left(\hat{x}_{t-1}, u_t\right)$ is a Jacobian matrix, which is the same as matrix $A$ in Kalman filtering we introduced at the beginning of the section. According to the same principle, based on our initial consideration of linear Kalman filtering in $\bf{Equation}$  $\bf{\ref{eq:kf}}$, the corresponding Jacobian matrix can be represented as:
\begin{equation}\begin{aligned} (A_{t-1})_{i j} & =\partial f_i / \partial x_j\left(\hat{x}_{t-1}, u_t\right) & (W_{t-1})_{i j} &=\partial f_i / \partial w_j\left(\hat{x}_{t-1}, u_t\right) \\ (H_t)_{i j} & =\partial h_i / \partial x_j\left(\hat{x}_k^{-}, u_t\right) & (\epsilon_t)_{i j} &=\partial h_i / \partial \epsilon_j\left(\hat{x}_k^{-}, 0\right)
\end{aligned}
\end{equation}

Then, under the same foundational assumptions guiding the Kalman Filter, we meticulously repeat the process:
\begin{algorithm}
\caption{Extended Kalman Filter}
\begin{algorithmic}[1]

\State \textbf{Input:} Initial state estimate $\hat{\mathbf{x}}_0$, error covariance $P_0$, system matrices $A_t$, $H_t$, $Q_t$, $R_t$, and measurement $y_t$.
\State \textbf{Output:} Updated state estimate $\hat{\mathbf{x}}_t$ and error covariance $P_t$.

\For{each time step $t = 1, 2, \ldots$}
    \State Prediction step:
    Propagate the a priori state estimate:
    $\hat{\mathbf{x}}_t^{-} = \mathbf{f}(\hat{\mathbf{x}}_{t-1}, 0)$
    
    \State Measurement prediction step:
    Produce the estimated measurement:
    $\hat{y}_t = h(\hat{\mathbf{x}}_t^{-}, 0)$
    
    \State Kalman gain computation:
    Compute the Kalman gain matrix:
    $K_t = P_t^{-} H_t^T (H_t P_t^{-} H_t^T + \epsilon_t R_t \epsilon_t^T)^{-1}$
    
    \State Update step:
    Correct the a priori estimate using the actual measurement:
    $\hat{\mathbf{x}}_t = \hat{\mathbf{x}}_t^{-} + K_t (y_t - \hat{y}_t)$
    
    \State Error covariance update:
    $P_t = (I - K_t H_t) P_t^{-}$ and $P_t^{-} = A_t P_{t-1} A_t^T + W_t Q_{t-1} W_t^T$
\EndFor

\end{algorithmic}
\end{algorithm}

Like what we introduced in Linear Kalman Regression. If we want to find the parameter, we have to construct the likelihood function with marginal likelihood function in $\bf{Equation}$ $\bf{\ref{eq:marginallikeihood}}$. By adjusting the equation with EKF, we have:
\begin{equation}\label{eq:likeEKF}
L_{1: T}=\sum_{t=1}^T\left[\ln \left(P_t\right)+\frac{{(y_t - \hat{x}_t^-})^2}{P_t}\right]
\end{equation}

\subsection{Particle Extended Kalman Filter (Particle-EKF)}
\label{sec:pekf}
So far, we introduced Kalman filters and EKF based on the assumption of Gaussian distribution. However, Gaussian distribution does not remain unchanged in nonlinearity\cite{GYORGY201465}. To address this issue, EKF focuses on mean and covariance. However, this method cannot consistently produce precise results, highlighting the need for more detailed solutions in such complex situations.

One method to solve this is by using a lot of sample points to show what the distribution looks like, which is known as particle filtering\cite{GYORGY201465}. The particle filter is based on Monte Carlo simulation. The use of particle sets to represent probabilities can be applied to any form of state space model. The core idea is to represent the distribution of random state particles extracted from a posterior probability, which is a sequential importance sampling method. Details about how to accomplish the particle EKF is presented in \textbf{Algorithm \ref{algo:ParticleEKF}}.

Particle filters do not necessarily assume Gaussian noise. Therefore, we do not need to maximize a Gaussian-assumed likelihood function for parameter estimation as we would with Linear Kalman Regression or the EKF. The text\cite{particle-liklihood} explains that the total likelihood of a given state, considering all past observations, is the product of various individual likelihoods. The logarithmic likelihood equation simplifies the process by converting multiplication into addition, thereby facilitating easier computation\cite{Tokdar2009}:
\begin{equation}\label{eq:particlelog}
\ln \left(L_{1: T}\right)=\sum_{t=1}^T \ln \left(l_t\right)= \sum_{t=1}^T \ln \left(\int p\left(y_t \mid x_t\right) \frac{p\left(x_t \mid z_{1: t-1}\right)}{q\left(x_t \mid x_{t-1}, y_{1: t}\right)} q\left(x_t \mid x_{t-1}, y_{1: t}\right) d x_t\right)
\end{equation}

The likelihood \(l_t\) for step \(t\) is approximated by a sum over \(N\) simulations. Each term in the sum is a ratio of the probability of observation \(y_t\) given the state \(x_t^{(i)}\), and a proposal distribution \(q\). This ratio is weighted by the probability of the state \(x_t^{(i)}\) given the previous state \(x_{t-1}^{(i)}\). This approximation is used for sequential importance sampling weight update in particle filtering algorithms.

\begin{equation}
l_t \approx \sum_{i=1}^{N} \frac{p\left(y_t \mid x_t^{(i)}\right) p\left(x_t^{(i)} \mid x_{t-1}^{(i)}\right)}{q\left(x_t^{(i)} \mid x_{t-1}^{(i)}, y_{1: t}\right)}
\end{equation}

\begin{algorithm}
\caption{\label{algo:ParticleEKF}Particle Extended Kalman Filter}
\begin{algorithmic}[1]

\State \textbf{Initialize:}
\State Set $t = 0$, simulation particles $N$, and choose initial state $x_0$ and initial covariance $P_0 > 0$
\For{$i = 1$ to $N$}
    \State Generate $x_0^{(i)} = x_0 + \sqrt{P_0} \cdot Z^{(i)}$ \Comment{$Z^{(i)}$ is a standard Gaussian random number}
    \State Set $P_0^{(i)}$ to $P_0$ and initial weight $w_0^{(i)} = 1/N$
\EndFor
\\
\State \textbf{Algorithm:}
\While{$t < N$}
    \For{each simulation index $i$}
        \State Apply Kalman Filter or  EKF to get $\hat{x}_t^{(i)}$ from $\hat{x}_{t-1}^{(i)}$
        \State $P_t^{(i)}$ is the associated a posteriori error covariance matrix.
        \State Generate $\tilde{x}_t^{(i)} = \hat{x}_t^{(i)} + \sqrt{P_t^{(i)}} \cdot Y^{(i)}$ \Comment{$Y^{(i)}$ is a standard Gaussian random number}
    \EndFor
    
    \For{each $i$ between 1 and $N$}
        \State Calculate the weights:
        \State $w_t^{(i)} = w_{t-1}^{(i)} \cdot \frac{p(y_t|\tilde{x}_t^{(i)}) \cdot p(\tilde{x}_t^{(i)}|\hat{x}_{t-1}^{(i)})}{q(\tilde{x}_t^{(i)}|\hat{x}_{t-1}^{(i)}, y_t)}$
        \Comment{$q()$ is the normal density with mean $\hat{x}_t^{(i)}$ and variance $P_t^{(i)}$}
    \EndFor
    
    \State Normalize the weights:
     $\hat{w}_t^{(i)} = w_t^{(i)} / \sum_{j=1}^{N} w_t^{(j)}$
    
    \State Resample the points $\tilde{x}_t^{(i)}$ and get $x_t^{(i)}$
    \State Reset weights $w_t^{(i)} = \hat{w}_t^{(i)} = 1/N$   
    \State t += 1
\EndWhile

\end{algorithmic}
\end{algorithm}

\section{SDE Simulation Results}

\subsection{Ornstein–Uhlenbeck (OU) Process}
The Ornstein-Unlenbeck Process is an SDE modeling interest rates. It is defined in the following formula:
\begin{equation} \label{eq:OUProcess}
    dX_t=\theta(\mu - X_t)dt + \sigma dW_t
\end{equation}
where $W_t$ is a Brownian motion. Parameters $\theta, \mu, \sigma$ are positive constants and to be estimated. Note that the $\theta$ here is not to be confused with $\theta$ in previous sections, which represents a general vector of parameters. The SDE could be solved analytically as: 
$X_t = X_0 e^{-\theta t} + \mu (1 - e^{-\theta t}) + \displaystyle\int_{0}^{t} \sigma e^{\theta(s-t)} dW_s$

\subsubsection{MLE of OU Process} \label{sec:MLE_OU}

Discretizing the SDE following $\bf{Equation}$ $\bf{\ref{eq:ItôDiffDisc}}$ gives:
$$X_t = X_{t-1} + \theta(\mu - X_{t-1}) dt + \sigma Z_{t-1}$$ where $dt$ is the time step, and $Z_{i}$ follows independently a normal distribution $N(0, dt)$

One computes the conditional density $f(X_t\vert X_{t-1}); \theta)$ using $\bf{Equation}$ $\bf{\ref{eq:ItoCondDensity}}$ and arrives at:

\begin{equation}
    f(X_t\vert X_{t-1}) = \frac{1}{\sqrt{2\pi dt}\sigma} \exp\left\{-\frac{1}{2}\frac{[(X_t - X_{t-1} - \theta(\mu - X_{t-1})]^2}{\sigma^2 \cdot dt}\right\}
\end{equation}

In numerical studies, we first fix a starting point $X_0$ of the SDE. We manually input parameters $\theta_0, \mu_0, \sigma_0$ and generate simulated data. Our time step is 0.499, and we propagated the process for 1000 iterations. Then, based on the simulated data, we optimize the log-likelihood function $\bf{Equation}$ $\bf{\ref{eq:LogLikelihood}}$ over choices of $\theta, \mu, \sigma$ using the above calculations with scientific computing software. We aim at accurately reconstructing the original parameters. Afterwards, we use the estimated parameters to reconstructed another set of data. Details are available in our code on GitHub.

Our original parameters are: $\theta = 1, \mu = 2, \sigma = 3$. The reconstructed parameters are: $\theta = 1.006, \mu = 2.065, \sigma = 3.187$ with running time: 4.418s. Plotting the original and reconstructed processes yields:

\begin{figure}[h!]
\begin{center}
\includegraphics[scale=.45]{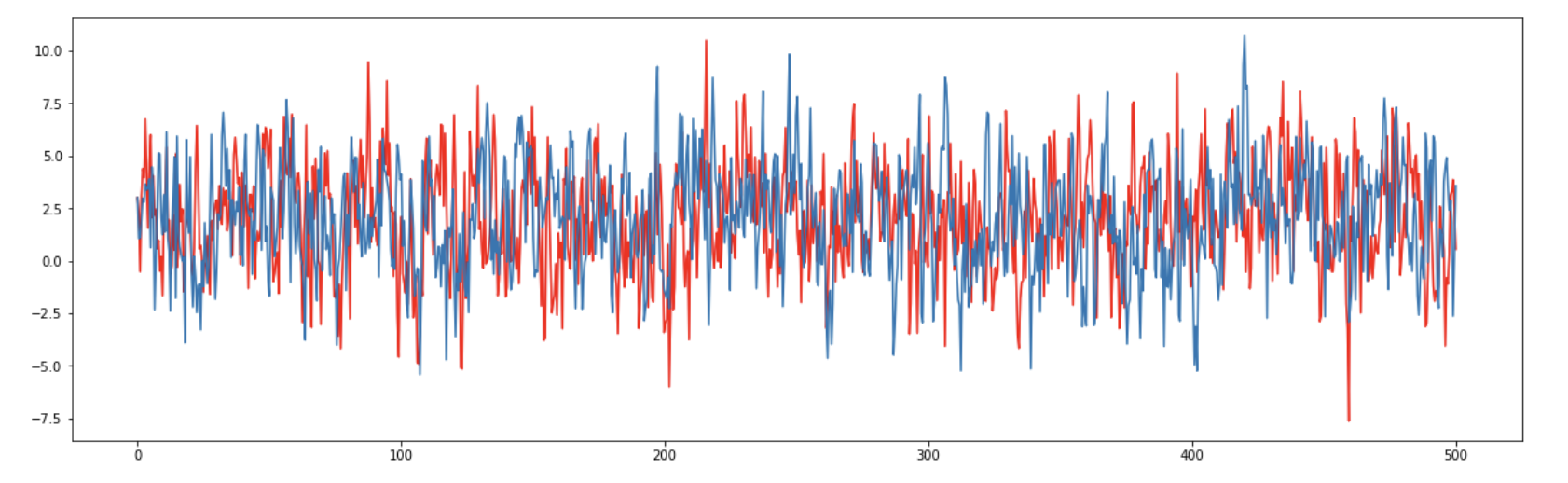}
\caption{Simulation of Ornstein–Uhlenbeck Process with original and MLE} 
\end{center}
\end{figure}

In the figure, the blue line is the original simulated data, and the red line is the reconstructed data. The horizontal axis is time, the vertical axis is the value of $X_t$.

We also applied MLE to estimate the Black–Karasinski (B-K) model, which is mainly used for the pricing of exotic interest rate derivatives such as American and Bermudan bond options. The SDE of the B-K model is given by:

\begin{equation} \label{eq:B-K Model}
d \ln(r) = ( \theta - \alpha \ln(r) ) dt + \sigma dW_t
\end{equation}

Similarly, using $\bf{Equation}$ $\bf{\ref{eq:ItoCondDensity}}$, the likelihood function of B-K model can be written as:

\begin{equation}
    f(\ln (r_t) \vert \ln (r_{t-1})) = \frac{1}{\sqrt{2\pi dt}\sigma} \exp\left\{-\frac{1}{2}\frac{[(\ln (r_t) - \ln (r_{t-1}) - ( \theta - \alpha \ln (r_t)) dt]^2}{\sigma^2 \cdot dt}\right\}
\end{equation}

Our original parameters are set to $\theta = 1.0$, $\alpha = 0.8$, $\sigma = 0.6$, $dt = 1$. MLE estimates the reconstructed parameters as: $\theta = 0.992$, $\alpha = 0.801$, $\sigma = 0.587$. Plotting the original and reconstructed processes yields:
\begin{figure}[h!]
\begin{center}
\includegraphics[scale=.4]{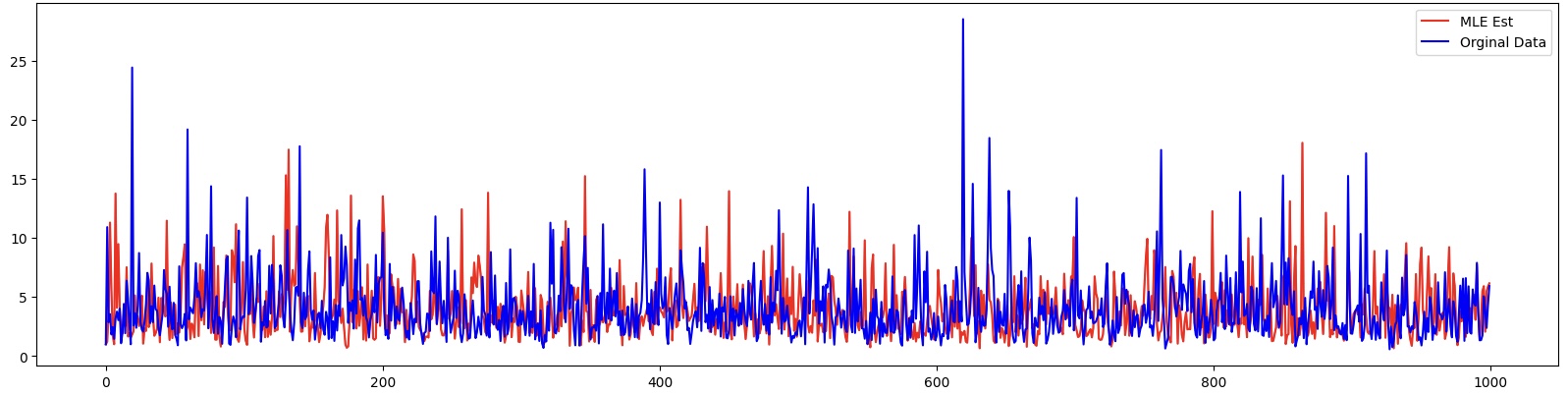}
\caption{Simulation of Black–Karasinski Model with original and MLE} 
\end{center}
\end{figure}


\subsubsection{Kalman Filter of OU Process}\label{sec:KF_OU}
Following linear system we introduced in Linear Kalman Filter, $\bf{Equation}$ $\bf{\ref{eq:LKRE}}$, the State and Measurement equations are:
\begin{equation}\label{eq:KF-OU}
\begin{aligned}
x_t &= \alpha+\beta x_{t-1}+G_tW_t \quad \text { with } \quad G_tW_t \sim \mathcal{N}\left(0, \sigma_{GW}^2\right)\\
y_t &= x_t+\epsilon_t \quad \text { with } \quad \epsilon_t \sim \mathcal{N}\left(0, \sigma_\epsilon^2\right)
\end{aligned}
\end{equation}
By the defined formula of OU process shows above, setting: $\alpha = \theta \mu  dt,\text{   }
    \beta = 1 - \theta dt, \text{   } 
    \text{var}[GW] = \sigma^2 dt, \text{  and  } 
    \text{var}[\epsilon] = 1 \times 10^{-4}$. And following $\bf{Equation}$ $\bf{\ref{eq:LinearKalm}}$, the new State and Measurement equations are:
\begin{equation}
\begin{aligned}
\left(\begin{array}{c}
1 \\
x_t
\end{array}\right) &= A_t\left(\begin{array}{c}
1 \\
x_{t-1}
\end{array}\right)+\left(\begin{array}{c}
0 \\
G_tW_t
\end{array}\right) \text { with } A_t = \left(\begin{array}{cc}
1 & 0 \\
\alpha & \beta
\end{array}\right), \text{ and } G_tW_t \sim \mathcal{N}\left(0, \sigma_{GW}^2\right)\\
y_t &=H_t\left(\begin{array}{c}
1 \\
x_{t}
\end{array}\right)+\epsilon_t \quad \text{ with } \quad H_t = 1 \text{ and } \epsilon_t \sim \mathcal{N}\left(0, \sigma_\epsilon^2\right)
\end{aligned}
\end{equation}

Predict steps:
\begin{equation}
\hat{x}_{t \mid t-1}=\alpha+\beta \hat{x}_{t-1 \mid t-1} \quad \text { and } \quad P_{t \mid t-1}=\beta^2 P_{t-1 \mid t-1}+\sigma_{GW}^2 \text {. }
\end{equation}

Supporting Variables:
\begin{equation}
\begin{aligned}r_k  = y_t-\hat{x}_{t \mid t-1}, \quad
S_t =P_{t \mid t-1}+\sigma_\epsilon^2, \quad
K_t =\frac{P_{t \mid t-1}}{S_t}
\end{aligned}
\end{equation}
Update step:
\begin{equation}
\hat{x}_{t \mid t}=\hat{x}_{t \mid t-1}+K_t r_t \quad \text { and } \quad P_{t \mid t}=P_{t \mid t-1}\left(1-K_t\right)
\end{equation}
In each iteration, the log likelihood function is calculated in $\bf{Equation}$ $\bf{\ref{eq:logLKR}}$.

\begin{figure}[h!]
\begin{center}
\includegraphics[scale=.45]{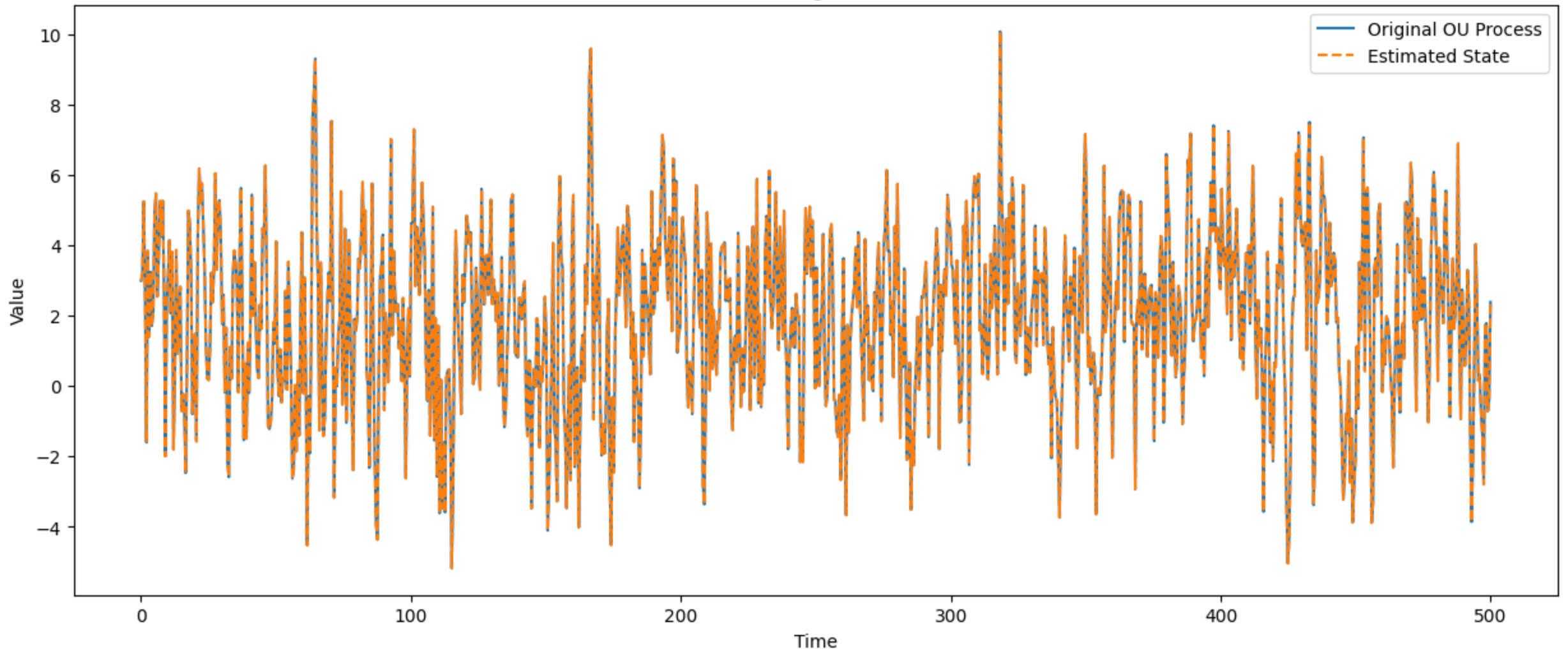}
\caption{\label{fig:figure2}OU Process Tracking with Kalman Filter} 
\end{center}
\end{figure}

$\bf{Figure}$ $\bf{\ref{fig:figure2}}$ shows the tracking normal OU process with Kalman Filter with same guesses parameter as MLE. The root mean square error (RMSE) between the two states is 4.560e-05, which performs a good fitting.

By minimizing the log-likelihood function with Limited-memory BFGS, the reconstructed parameters are: $\theta = 0.9267$, $\mu = 1.996$, $\sigma = 2.902$ with the original parameters are as same as the previous OU Process part ( $\theta = 1, \mu = 2, \sigma = 3$). In terms of estimating the parameters, there is no significant difference between the two results, but the parameter estimation method using Kalman filtering only runs for 0.30717s.

\subsection{OU Process with Jump}\label{sec:OU-Jump}
For application purposes, it is often realistic to add a jump process in addition to Brownian noise. As an example, the discretized OU process can be adapted to
\begin{equation} \label{eq:OU_Disc_Jump}
    X_t = X_{t-1} + \theta(\mu - X_{t-1}) dt + \sigma Z_{t-1} + J_{t-1}
\end{equation}
where each $J_{i}$ takes parameters $\lambda_j, \mu_j, \sigma_j$ and behaves identically and independently in the following way:

\begin{enumerate}
    \item sample a random number $\gamma \sim N(0, 1)$
    \item if $\gamma < \lambda_j$, $J_i$ is "active", i.e., sample $J_i \sim N(\mu_j, \sigma_{j}^2)$
    \item if $\gamma \ge \lambda_j$, $J_i = 0$
\end{enumerate}

We remark that generally jump processes could be much more complicated.

\subsubsection{MLE of OU Process with Jump}
With the above example, one can in fact still analytically compute the log-likelihood function. This results from the design of $J_i$ by our assumption.

Following our previously introduced framework, we focus on computing the conditional density $f(X_t \vert X_{t-1})$. This is done by considering the two cases described in bullet points 2 and 3 above. If $J_{t-1}$ is active (case 2), $X_t$ follows a normal distribution centered at $\left(X_{t-1} + \theta (\mu-X_{t-1})dt + \mu_j\right)$, with variance $\sigma^2 dt + \sigma_j^2$. If $J_{t-1}$ is identically zero (case 3), $X_t \sim N\left(X_{t-1} + \theta (\mu-X_{t-1})dt, \sigma^2 dt\right)$ as it is in the simple OU process.

With this established, knowing the density for normal distributions, we compute:
\begin{align*}
    P(X_t \vert X_{t-1}) = &= P(X_{i+1}, J_i = 0| X_i) + P(X_{i+1}, J_i \ne 0| X_i) \\
    &= P(X_{i+1} | J_i = 0, X_i) \cdot P(J_i = 0) + P(X_{i+1} | J_i \ne 0, X_i)\cdot P(J_i \ne 0) \\
    &= (1-F_N(\lambda_j dt))\cdot \frac{1}{\sqrt{2\pi dt}\sigma} \cdot\exp\left\{-\frac{1}{2}\left(\frac{X_{i+1}-[X_{i}+\theta(\mu-X_i)dt]}{\sigma\sqrt{dt}}\right)^2\right\} dx \\
    &+ F_N(\lambda_j dt)\cdot \frac{1}{\sqrt{2\pi(\sigma^2 dt + \sigma_j^2)}} \cdot \exp \left\{ -\frac{1}{2} \left(\frac{X_{i+1} - [X_{i}+\theta(\mu-X_i)dt + \mu_j]}{\sqrt{\sigma^2 dt + \sigma_j^2}}\right)^2 \right\} dx
\end{align*}
where we used Bayes' Theorem in the second step. $F_N$ denotes the cumulative distribution function for the standard normal distribution, and the $dx$ is the technicality term for calculating probability. Omitting the $dx$ gives the conditional density $f(X_t \vert X_{t-1})$. 

\begin{figure}[h!]
\begin{center}
\includegraphics[scale=.4]{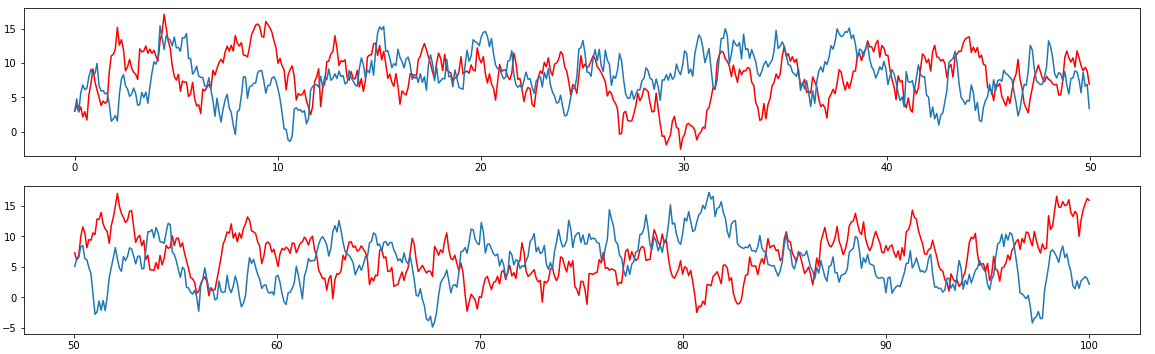}
\caption{\label{fig:MLE_OU_J5}Simulation of Ornstein–Uhlenbeck Process with jump and MLE} 
\end{center}
\end{figure}

With such, we repeated the procedure in $\bf{Section}$ $\bf{\ref{sec:MLE_OU}}$ to examine the effectiveness of MLE in this setting. We enforced a bound of $(10^{-15}, 6)$ for all parameters during optimization. The results are summarized in the following  with running time: 27.83828s:

\begin{table}[ht]
    \centering
    \begin{tabular}{|c|c|c|c|c|c|c|}
        \hline 
        Parameter & $\theta$ & $\mu$ & $\sigma$ & $\lambda_j$ & $\mu_j$ & $\sigma_j$ \\
        \hline 
        Original & 1 & 2 & 4 & 0.5 & 1 & 1 \\
        \hline 
        Reconstructed & 0.86 & 0 & 4.005 & 0 & 1.221 & 0.798 \\
        \hline
    \end{tabular}
    \vspace{0.25em}
    \caption{MLE results for OU process with jump}
    \label{tab:MLE_OU_Jump}
\end{table}
In the following \textbf{Figure \ref{fig:MLE_OU_J5}}, the blue lines represent original simulated data, and the red lines are reconstructed data.

\subsubsection{Kalman Filter of OU Process with Jump}
In the example of the OU process with a jump, given its linear nature, the Linear Kalman Filter can be effectively employed to describe the state and measurement, as detailed in $\bf{Equation}$ $\bf{\ref{eq:KF-OU}}$. The thing we need to aware of that adding jump term to the noise, which means: $\text{var}[GW] = \sigma_{GW}^2 dt + \lambda_j \cdot \mu_j^2dt$. The remaining parts of the equation, including the prediction and update steps, follow the methodology outlined in $\bf{Section}$ $\bf{\ref{sec:KF_OU}}$.

\begin{figure}[h!]
\begin{center}
\includegraphics[scale=.22]{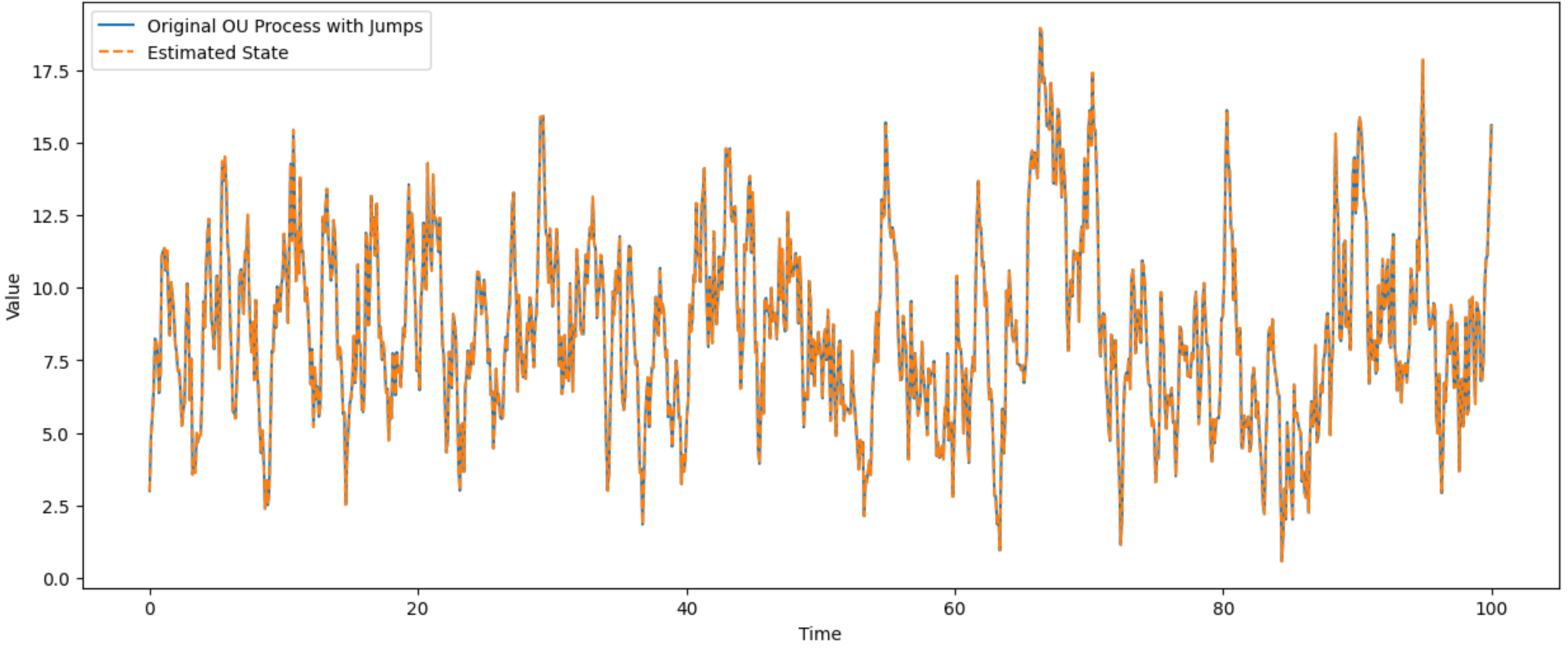}
\caption{\label{fig:figure3} OU Process with Jump Tracking with Kalman Filter} 
\end{center}
\end{figure}

The illustration in $\bf{Figure}$ $\bf{\ref{fig:figure3}}$ depicts the tracking of an OU process that includes a jump term, utilizing a Kalman Filter with parameters identical to those used in the MLE. The RMSE between the two states is 0.0001021, indicating a high degree of accuracy in fitting.

By minimizing the log-likelihood function using Limited-memory BFGS, the reconstructed parameters are obtained as:

\begin{table}[ht]
    \centering
    \begin{tabular}{|c|c|c|c|c|c|c|}
        \hline
        Parameter & $\theta$ & $\mu$ & $\sigma$ & $\lambda_j$ & $\mu_j$ & $\sigma_j$ \\
        \hline 
        Original & 1 & 2 & 4 & 0.5 & 1 & 1 \\
        \hline 
        Reconstructed & 0.7858 & 6.0 & 4.844 & 0.5525 & 1.496 & 1.0 \\
        \hline
    \end{tabular}
    \vspace{0.25em}
    \caption{Kalman filter results for OU process with jump}
    \label{tab:MLE_OU_Jump}
\end{table}

We found the results of the $\mu$ term to be inaccurate. This could be attributed to the filters' comparative lack of sensitivity towards the parameter set, or perhaps due to the non-uniqueness of the optimal parameter configuration. 

The method of parameter estimation by using the solution of the Kalman filter does not show better results very intuitively compared to the MLE. However, compared to the 27.83828s spent by MLE, the running time of Kalman filter is 0.265414s. There is a significant difference between them.

Next, we aim to evaluate a more comprehensive model within our Kalman Filter framework to ascertain its efficacy in more complex scenarios.

\subsection{Heston Model}
In 1993, Heston enhanced the traditional Black-Scholes (BS) model by introducing stochastic volatility\cite{heston}, thereby deriving a model tailored for European option pricing. Heston formulated the SDE under the Physical Measure as:
\begin{equation}\label{eq:heston}
\begin{gathered}
d S_t=\mu S_t d t+\sqrt{v_t} S_t d W_{t}^1, \\ 
d v_t=\kappa\left(\theta-v_t\right) d t+\xi \sqrt{v_t} d W_{t}^2, \\ 
d W_{1} \cdot d W_{2}=\rho d t 
\end{gathered}
\end{equation}
In $\bf{Equation}$ $\bf{\ref{eq:heston}}$, $S_t$ is the price of the asset, $v_t$ is the instantaneous variance from the CIR process. $W_{t}^1$ and $W_{t}^2$ are two wiener process with instantaneous correlation $\rho$. The parameter $\mu$ is the risk-neutral rate of return, $\kappa$ is the adjustment rate, $\theta$ is the long-term run average variance, and $\xi$ is the volatility fluctuations \cite{MrázekPospíšil+2017+679+704}.

By expressing \( dS_t \) as \( d\ln(S_t) \),
this form simplifies the calculation of volatility, which typically impacts the rate of return rather than the absolute price. After transforming by Ito's Lemma, we get:

\begin{equation}
d \ln \left(S_t\right)=\left(\mu-\frac{1}{2} v_t\right) d t+\sqrt{v_t} d W_t^1
\end{equation}

Since volatility cannot be negative, we use the Feller condition: $\frac{2 \kappa \theta}{\xi^2}>1$ \cite{Feller}. Also, since pricing needs to be guaranteed to be arbitrage-free, we need to convert the SDE under Physical Measure to Martingale Measure, using the market price of risk:
\begin{equation}
\begin{aligned}
d \ln \left(S_t\right) = \left(\mu_s-\frac{1}{2} v_t\right) d t+\sqrt{v_t}  \widetilde{W}_{t}^1, \qquad  \widetilde{W}_{t}^1 &= d W_{t}^1+\frac{\mu-\mu_s}{\sqrt{v_t}} d t\\ 
d v_t = [\kappa\left(\theta-v_t\right) - \lambda v(t)] d t+\xi \sqrt{v_t}  \widetilde{W}_{t}^2, \qquad
 \widetilde{W}_{t}^2 &= d W_{t}^2+\frac{\left(W_t^2-\rho W_t^1\right)}{\sqrt{1-\rho^2}}d t
\end{aligned}
\end{equation}

then:
\begin{equation}
v_t = v_{t-1}+\left(\rho \xi-\theta v_{t-1}\right) d t+\xi \sqrt{v_{t-1}} \sqrt{d t} W_{t-1}^2 -\rho \xi\left[\ln S_{t-1}+\left(\mu_s-\frac{1}{2} v_{t-1}\right) d t+\sqrt{v_{t-1}} \sqrt{d t} W_{t-1}^1-\ln S_t\right]
\end{equation}

Considering the uncorrelated with $W_t^1$ and $W_t^2$, we have the observe and measurement for Heston:
\begin{equation}
\begin{aligned}
x_t &= v_t=v_{t-1}+\left[\left(\kappa\left(\theta-v_t\right)-\rho \xi \mu_s\right)-\left(\theta-\frac{\rho \xi}{2}\right) v_{t-1}\right] d t +\rho \xi \ln \left(\frac{S_t}{S_{t-1}}\right)+\xi \sqrt{1-\rho^2} \sqrt{v_{t-1}} \sqrt{d t} \tilde{W}_{t-1}^2 \\
y_t &= \ln S_t = \ln S_{t-1}+\left(\mu-\frac{1}{2} v_{t-1}\right) d t+\sqrt{v_{t-1}} \sqrt{dt} W_{t-1}^1\
\end{aligned}
\end{equation}

Overall, the Heston model is a two-factor nonlinear model. We will fit it by using filtering. After that, based on our filtering result, we will try to find the parameters based on minimizing the log-likelihood function. 

\begin{figure}[h!]
\begin{center}
\includegraphics[scale=.45]{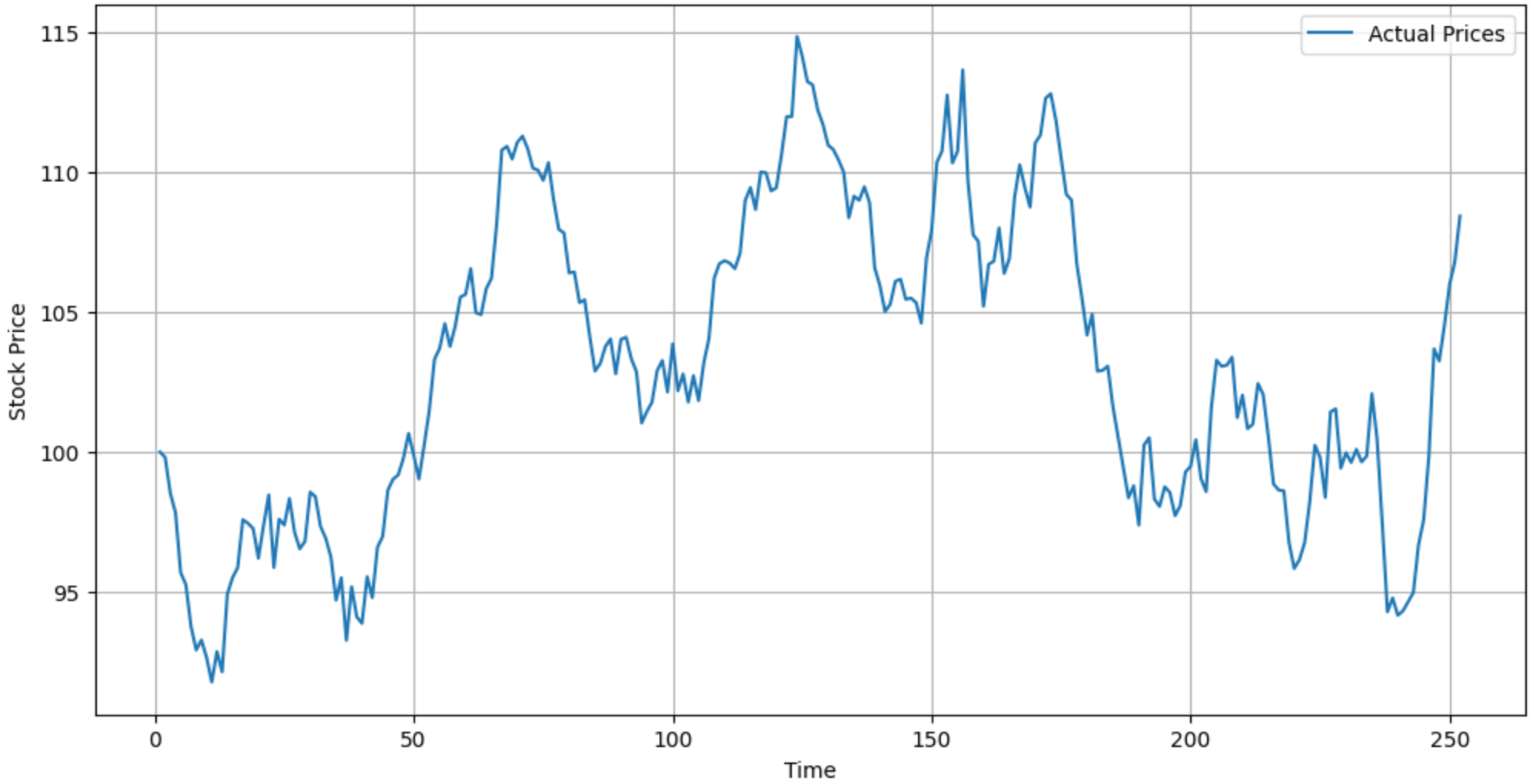}
\caption{\label{fig:Heston} Heston with Euler–Maruyama Method} 
\end{center}
\end{figure}

\subsubsection{Extended Kalman Filter in Heston}

In terms of the EKF, the following state equation is presented in Heston model:

\begin{equation}
x_t=f\left(x_{t-1}, w_{t-1}\right)=\left(\begin{array}{c}
\ln S_{t-1}+\left(\mu_S-\frac{1}{2} v_{t-1}\right) dt+\sqrt{v_{t-1}} \sqrt{dt} W_{t-1}^1 \\
v_{t-1}+\left(\kappa\left(\theta-v_t\right) -\theta v_{t-1}\right) dt+\xi \sqrt{v_{t-1}} \sqrt{dt} W_{t-1}^2
\end{array}\right)
\end{equation}

Following the procedure outlined shown in the input part of $\textbf{Algorithm 2}$, the pertinent matrix is displayed below:
\begin{equation}
Q_t=\left(\begin{array}{ll}
1 & \rho \\
\rho & 1
\end{array}\right), \quad A_t=\left(\begin{array}{cc}
1 & -\frac{1}{2} dt \\
0 & 1-\theta dt
\end{array}\right), \quad W_t=\left(\begin{array}{cc}
\sqrt{v_{t-1}} \sqrt{dt} & 0 \\
0 & \xi \sqrt{v_{t-1}} \sqrt{dt}
\end{array}\right), \quad w_t=\left(\begin{array}{c}
W_t^1 \\
W_t^2
\end{array}\right)
\end{equation}

We follow the formulation by Alireza Javaheri\cite{Inside-Volatility-Arbitrage} and write state-space into one dimension:
\begin{equation}
A_t=1-\left(\theta-\frac{1}{2} \rho \xi\right) d t,
\quad
W_t=\xi \sqrt{1-\rho^2} \sqrt{v_{k-1}} \sqrt{d t},
\quad
H_t=-\frac{1}{2} d t,
\quad
Q_t=\sqrt{v_t} \sqrt{d t}
\end{equation}

\begin{figure}[h!]
\begin{center}
\includegraphics[scale=.45]{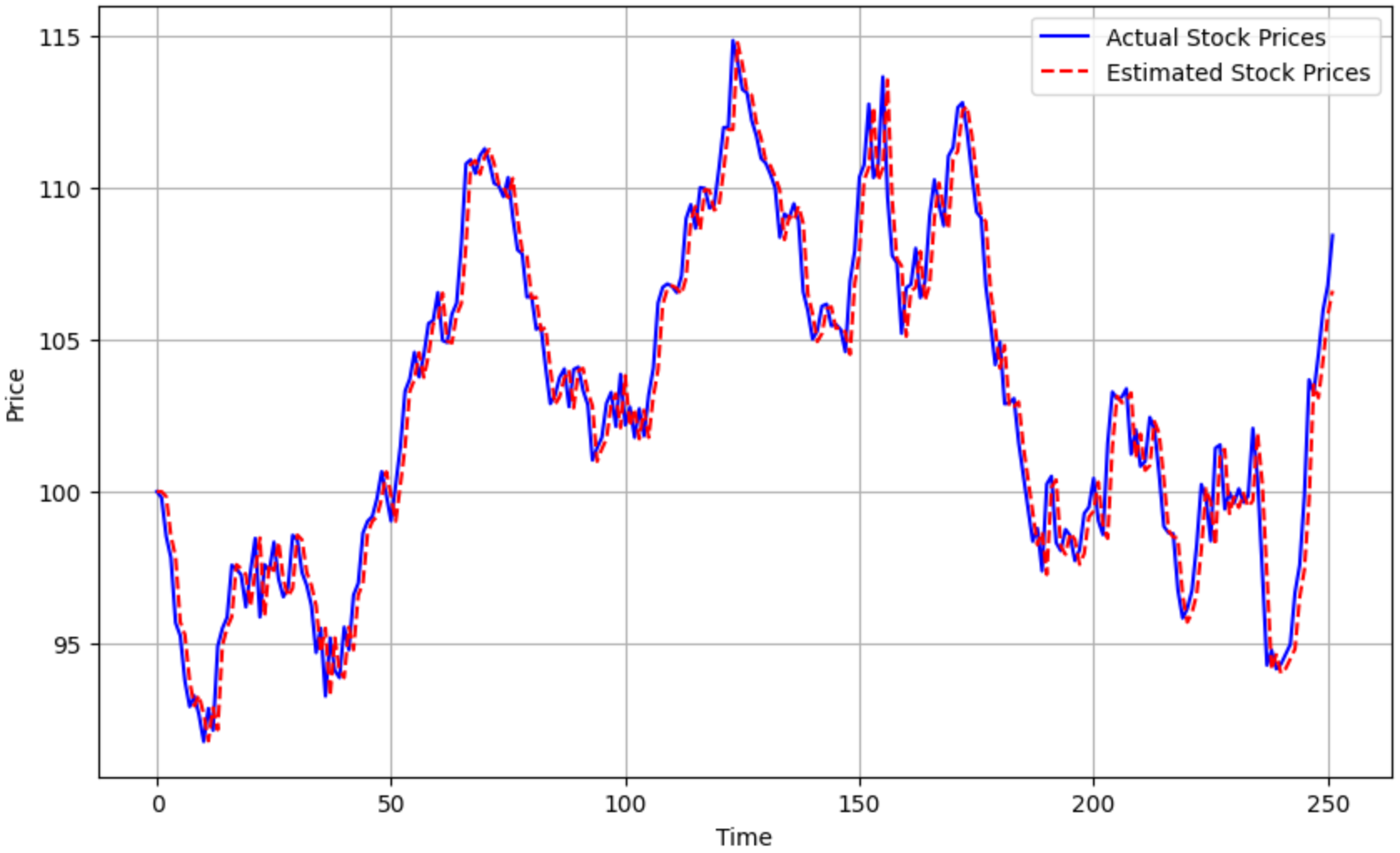}
\caption{\label{fig:Heston-EKF} Heston Tracking with EKF} 
\end{center}
\end{figure}

The illustration in \textbf{Figure} \ref{fig:Heston-EKF} depicts the tracking of a Heston model using an EKF with parameters identical to those in the likelihood function. The RMSE between the two states is 1.268415 and running time 0.01886s, indicating a high degree of accuracy. Regarding parameter estimation, by minimizing the log-likelihood function as shown in \textbf{Equation} \ref{eq:particlelog}, we obtained the estimated parameters presented in \textbf{Table} \ref{tab:EKF_Heston}.

\begin{table}[ht]
    \centering
    \begin{tabular}{|c|c|c|c|c|c|}
    \hline & $\boldsymbol{\mu}$ & $\boldsymbol{\theta}$ & $\boldsymbol{\rho}$ & $\boldsymbol{K}$ & $\boldsymbol{\xi}$ \\
    \hline Actual & 0.05 & 1.5 & 0.04 & 0.3 & -0.6 \\
    \hline Guess & 0.1 & 1.0 & 0 & 0.1 & 0 \\
    \hline Estimate & 1 & 2 & 0 & 0.5 & -1 \\
    \hline
    \end{tabular}
    \vspace{0.25em}
    \caption{EKF Estimate Heston's parameters}
    \label{tab:EKF_Heston}
\end{table}

We found the results for the $\mu$ term to be inaccurate, though other predictions appear reasonable. Reasonable estimates of parameters and low RMSE values suggest that the application of the EKF in the Heston model has been successful. 

\subsubsection{Particle Extended Kalman Filter in Heston}
In \textbf{Section 4.3}, we introduced the Particle Extended Kalman Filter. In this section, we adapt it to a one-dimensional state space with the following formulation:
\begin{equation}
\begin{aligned}
n(x, m, s) &= \frac{1}{\sqrt{2 \pi} s} \exp \left(-\frac{(x-m)^2}{2 s^2}\right), \
q\left(\tilde{v}t^{(i)} \mid x{t-1}^{(i)}, \ln S_{0: t-1}\right) &=n\left(\tilde{x}_t^{(i)}, m=\hat{x}_t^{(i)}, s=\sqrt{P_t^{(i)}}\right),
\end{aligned}
\end{equation}
where the normal density is characterized by mean $m$ and standard deviation $s$. Consequently, we derive:
\begin{equation}\label{eq:42}
p\left(\ln S_{t-1} \mid \tilde{x}k^{(i)}\right) = n\left(\ln S{t-1}, m=\operatorname{Ln} S_t+\left(\mu_S-\frac{1}{2} \tilde{x}_t^{(i)}\right) dt, s=\sqrt{\tilde{x}_t^{(i)}} \sqrt{dt}\right),
\end{equation}
and:
\begin{equation}\label{eq:43}
\begin{aligned}
p\left(\tilde{x}t^{(i)} \mid x{t-1}^{(i)}\right) = n\left(\tilde{x}t^{(i)}, m=x{t-1}^{(i)}+\left(\kappa\left(\theta-x_{t-1}\right)-\rho \xi \mu_S\right)dt -\frac{1}{2} \rho \xi x_{t-1}^{(i)}dt+\rho \xi\left(\ln S_t-\ln S_{t-1}\right), s\right).
\end{aligned}
\end{equation}
This provides us with the necessary densities for the filter implementation:
\begin{equation}
s = \xi \sqrt{1 - \rho^2} \sqrt{x_{t-1}^{(i)}} \sqrt{dt}
\end{equation}
With 1000 particles being used, the estimation of the observed state is given by:
\begin{equation}
\ln S_{t-1}^{-} = \frac{1}{1000} \sum_{i=1}^{1000} \ln \hat{S}_{t-1}^{(i)}
\end{equation}
obtained from the EKF applied to $x_{t-1}$. See $\bf{Algorithm 3}$ for more details.

\begin{figure}[h!]
\begin{center}
\includegraphics[scale=.45]{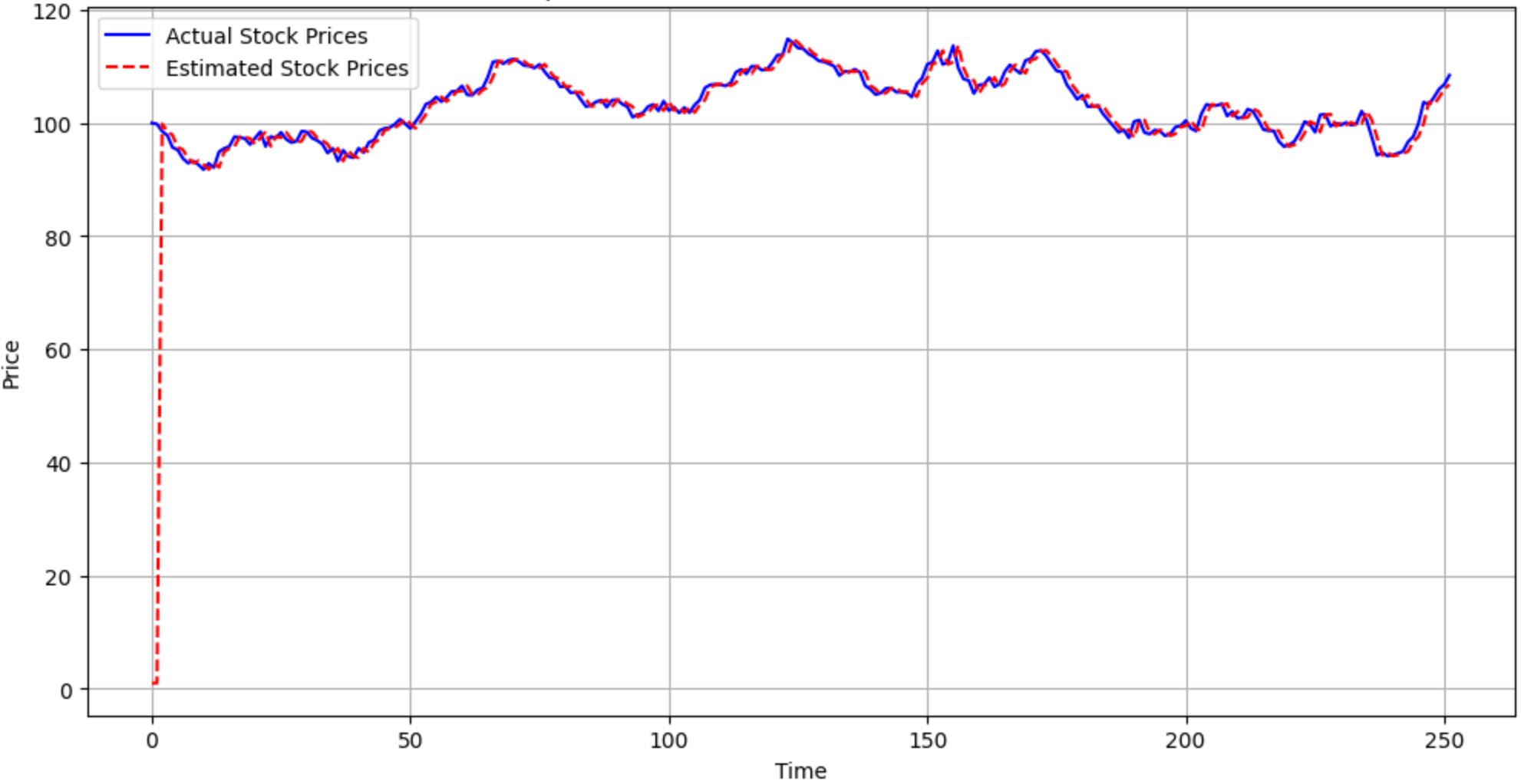}
\caption{\label{fig:HestonParticleEKF} Heston Tracking with Particle-EKF} 
\end{center}
\end{figure}

\textbf{Figure} \ref{fig:HestonParticleEKF}, illustrates the tracking performance of a Heston model using the Particle-EKF. With an RMSE score of 8.9011336, running time of 130.56615s, the results indicate a commendable level of fit.

\subsection{Bates Model}
The bates model is based on the Heston model, which extends the volatility by incorporating a Merton log-normal jump\cite{jumps}. As $\bf{Equation}$ $\bf{\ref{eq:bates}}$ shows below, consequently, the number of jumps occurring in the interval between $t+dt$ will be denoted as $dN_t$, which is a Poisson counter.
\begin{equation}\label{eq:bates}
d S_t=\left(\mu_s+\lambda j\right) S_t d t+\sqrt{v_t} S_t d W_t-S_t j d N_t
\end{equation}
Here, $\lambda$ is intensity, and j is the Gaussian random variable that has a fixed fractional jump size $[0, 1)$, which has a close definition in $\bf{Section}$ $\bf{\ref{sec:OU-Jump}}$. And by applying Ito's lemma for semi-Martingales:
\begin{equation}
d \ln S_t=\left(\mu_s-\frac{1}{2} v_t+\lambda j\right) d t+\sqrt{v_t} d W_t+\ln (1-j) d N_t
\end{equation}

\begin{figure}[h!]
\begin{center}
\includegraphics[scale=.45]{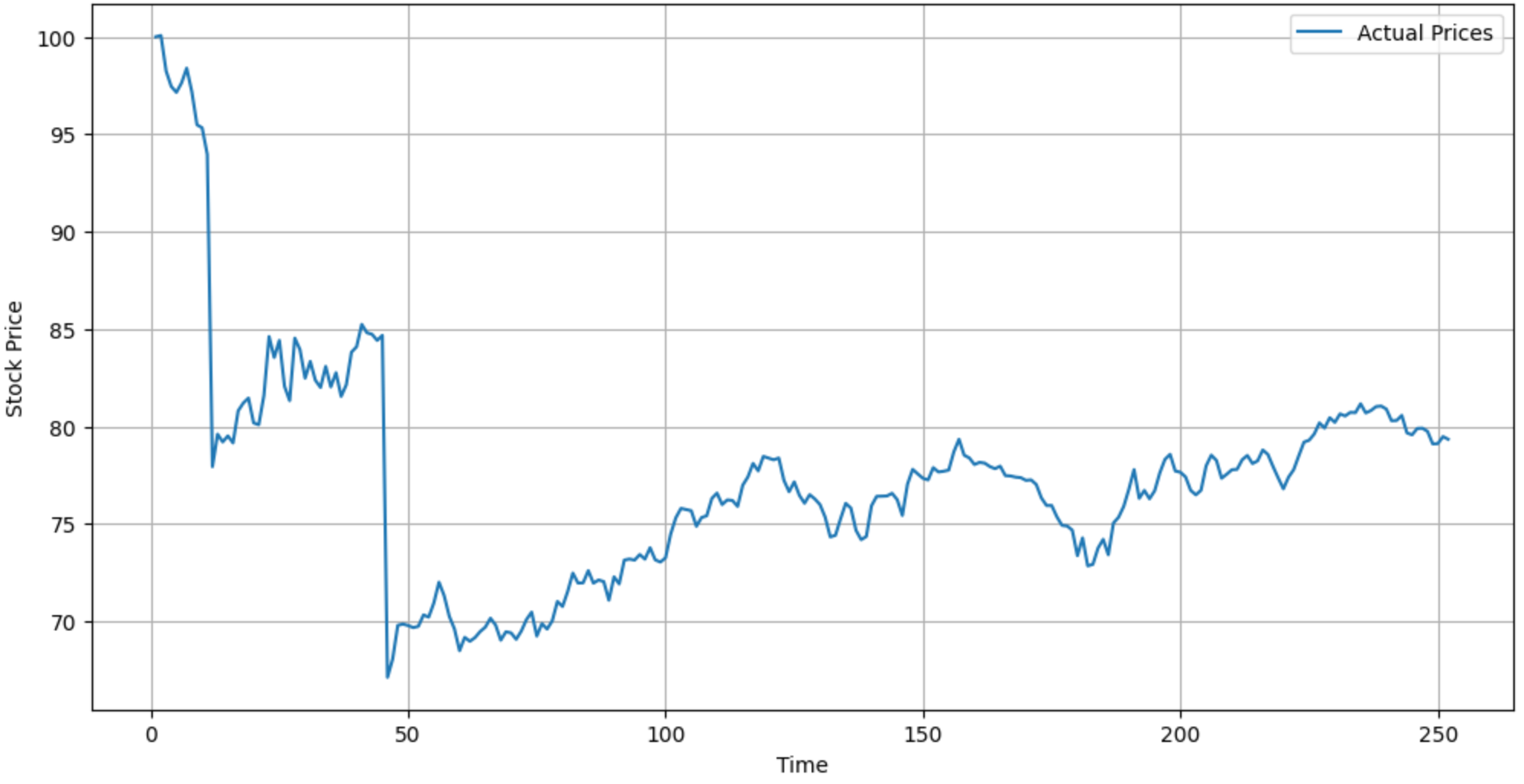}
\caption{\label{fig:Bates} Bates with Euler–Maruyama Method} 
\end{center}
\end{figure}

As same as what we have down in Heston, we will fit this model by using filtering. Then, we have the transition matrix:
\begin{equation}
x_t=\left(\begin{array}{c}\delta_0(0) e^{-\lambda d t}+\delta_0(\ln (1-j))\left(1-e^{-\lambda d t}\right) \\ v_{k-1}+\left[\left(\kappa\left(\theta-v_t\right)-\rho \xi\left(\mu_s+\lambda j\right)-\left(\theta-\frac{1}{2} \rho \xi\right) v_{t-1}\right] d t+\rho \xi\left[\ln \left(\frac{S_t}{S_{t-1}}\right)-\mu_{t-1}\right]+\xi \sqrt{1-\rho^2} \sqrt{v_{t-1}} \sqrt{dt} \tilde{W}_{t-1}^2\right.\end{array}\right)
\end{equation}

\subsubsection{Extended Kalman Filter in Bates}
We have imployed a similar method in $\bf{Section 5.2.2}$, with adapting EKF to processes with jumps, by adding $\lambda j$ to all $\mu_s$. Following the $\bf{Algorithm 2}$, we got:

\begin{figure}[h!]
\begin{center}
\includegraphics[scale=.45]{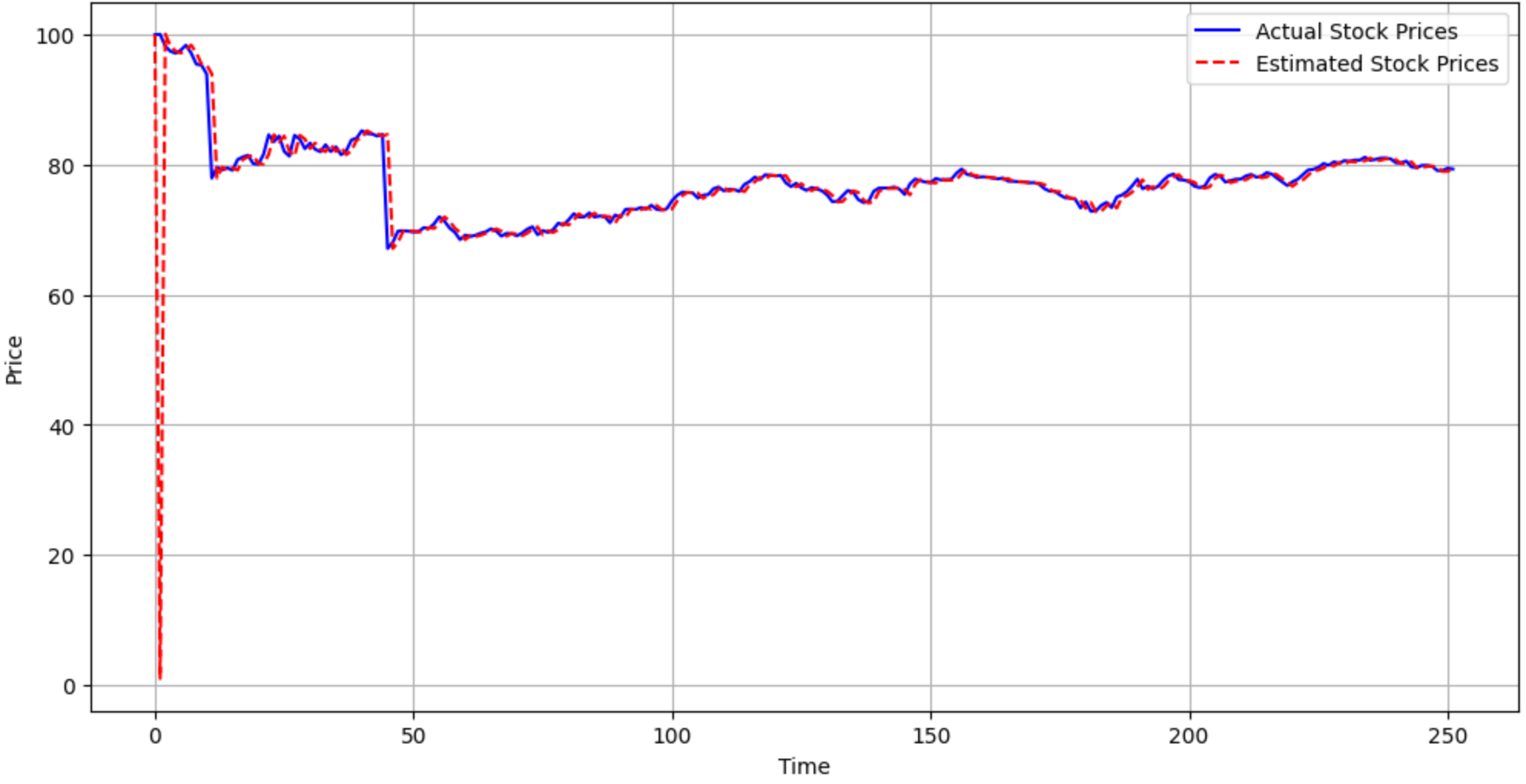}
\caption{\label{fig:BatesEKF} Bates Tracking with EKF} 
\end{center}
\end{figure}

The illustration in \textbf{Figure} \ref{fig:Heston-EKF} depicts the tracking of a Heston model using an EKF with parameters identical to those in the likelihood function. The RMSE between the two states is 6.37269 and running time is 0.0081274, which is not a bad solution. Regarding parameter estimation, by minimizing the log-likelihood function as shown in \textbf{Equation} \ref{eq:likeEKF}, we obtained the estimated parameters presented in \textbf{Table} \ref{tab:EKF_Bates}.

\begin{table}[ht]
    \centering
    \begin{tabular}{|c|c|c|c|c|c|c|c|}
    \hline & $\boldsymbol{\mu}$ & $\boldsymbol{\theta}$ & $\boldsymbol{\rho}$ & $\boldsymbol{K}$ & $\boldsymbol{\xi}$ & $\lambda$ & j \\
    \hline Actual & 0.05 & 1.5 & 0.04 & 0.3 & -0.6 & 10 & 0.1 \\
    \hline Guess & 0.1 & 1.0 & 0 & 0.1 & 0 & 8 & 0.1 \\
    \hline Estimate & 1 & 2 & 0 & 0.5 & -1 & 6 & 0.1 \\
    \hline
    \end{tabular}
    \vspace{0.25em}
    \caption{EKF Estimate Bates' parameters}
    \label{tab:EKF_Bates}
\end{table}

However, parameter estimation results yielded by EKF are not ideal. This could be attributed to the filters’ comparative lack of sensitivity towards the parameter set, or perhaps due to the non-uniqueness of the optimal parameter configuration.

\subsubsection{Particle Extended Kalman Filter in Bates}
The approach of using Particle-EKF in the Bates model is similar to our implementation in the Heston model. The adjustment required is the addition of $\lambda j$ to $\mu_s$. For instance, refer to $\bf{Equation}$$\bf{\ref{eq:42}}$ and $\bf{Equation}$$\bf{\ref{eq:43}}$:
\begin{equation}
\begin{aligned}
p\left(\ln S_{t-1} \mid \tilde{x}k^{(i)}\right) &= n\left(\ln S{t-1}, m=\operatorname{Ln} S_t+\left((\mu_s + \lambda j)-\frac{1}{2} \tilde{x}_t^{(i)}\right) dt, s=\sqrt{\tilde{x}_t^{(i)}} \sqrt{dt}\right),
\\
p\left(\tilde{x}t^{(i)} \mid x{t-1}^{(i)}\right) &= n\left(\tilde{x}t^{(i)}, m=x{t-1}^{(i)}+\left(\kappa\left(\theta-x_{t-1}\right)-\rho \xi (\mu_s + \lambda j)\right) dt -\frac{1}{2} \rho \xi x_{t-1}^{(i)} dt+\rho \xi\left(\ln S_t-\ln S_{t-1}\right), s\right).
\end{aligned}
\end{equation}

By simplifying these elements in one dimension form, we obtain:
\begin{equation}
\begin{aligned} 
p\left(ln S_{t+1} \mid \tilde{x}_k^{(i)}\right) & =n\left(ln S_{t+1}, ln S_{t}+\left(\mu_s+\lambda j-\frac{1}{2} \tilde{x}_t^{(i)}\right) d t+\tilde{\mu}_t^{(i)}, \sqrt{\tilde{x}_t^{(i)} d t}\right) \\ p\left(\tilde{x}_t^{(i)} \mid x_{t-1}^{(i)}\right) & =n\left(\tilde{x}_t^{(i)}, m, s=\xi \sqrt{1-\rho^2} \sqrt{x_{t-1}^{(i)}} \sqrt{d t}\right) p\left(\tilde{\mu}_t^{(i)} \mid \mu_{t-1}^{(i)}\right)
\end{aligned}
\end{equation}
And then, we have the mean and particles:
\begin{equation}
\begin{aligned}
{m} &= x_{t-1}^{(i)}+\left[\kappa\left(\theta-x_t\right)-\rho \xi\left(\mu_s+\lambda j\right)-\left(\theta-\frac{1}{2} \rho \xi\right) x_{t-1}^{(i)}\right] d t+\rho \xi \ln \left(\frac{S_k}{S_{k-1}}\right)-\rho \xi \mu_{t-1}^{(i)}
\\
q\left(\tilde{x}_t^{(i)} \mid x_{t-1}^{(i)}, ln S_{1: t+1}\right) &= n\left(\tilde{x}_t^{(i)}, \hat{x}_t^{(i)}, P_t^{(i)}\right) p\left(\tilde{\mu}_t^{(i)} \mid \mu_{t-1}^{(i)}\right)
\end{aligned}
\end{equation}
And following $\bf{Algorithm 3}$, we got:
\begin{figure}[h!]
\begin{center}
\includegraphics[scale=.45]{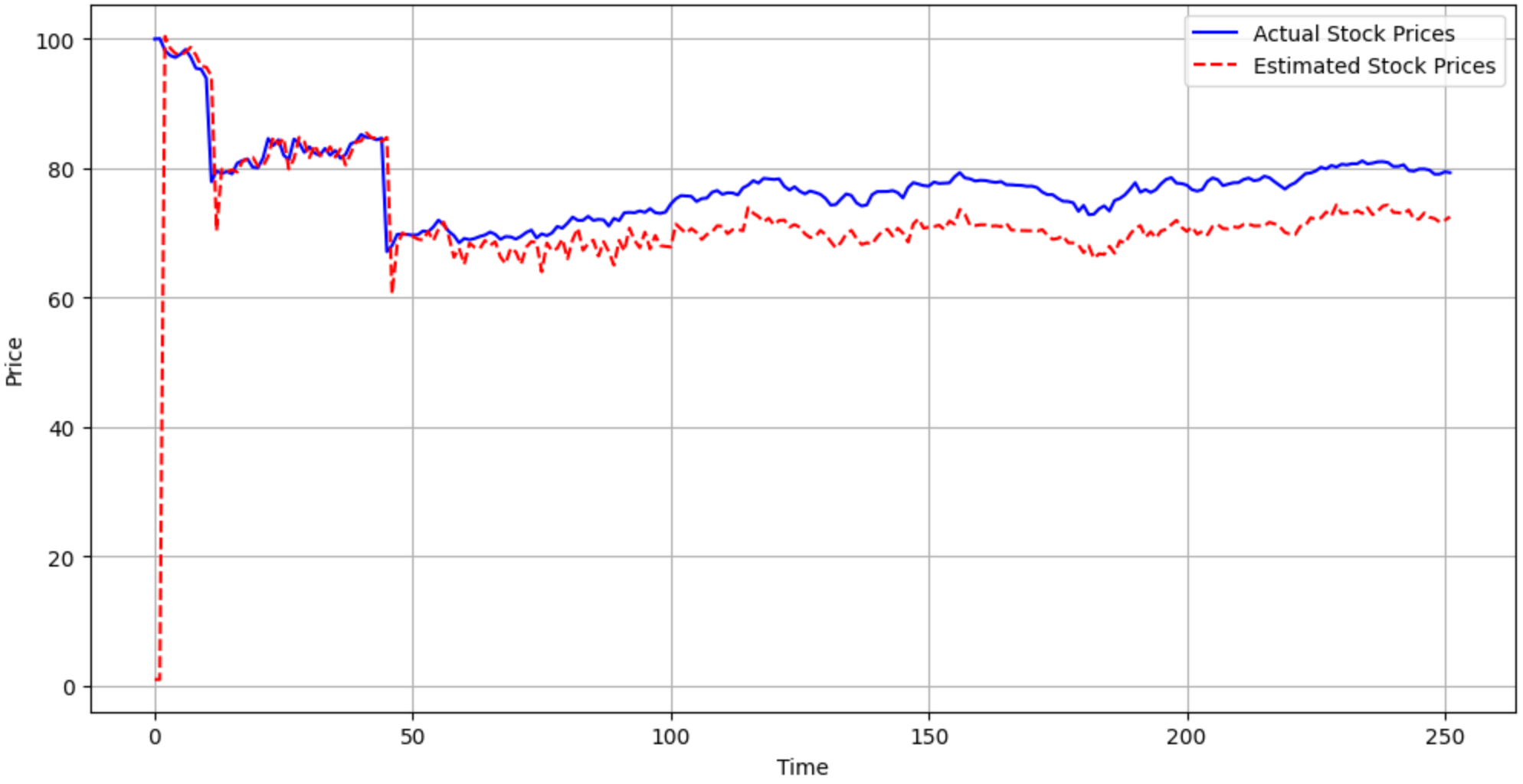}
\caption{\label{fig:BatesPart} Bates Tracking with Particle-EKF} 
\end{center}
\end{figure}

The figure referenced above, \textbf{Figure} \ref{fig:BatesPart}, illustrates the tracking performance of a Bates model using Particle-EKF. With an RMSE score of 10.48056 and running time 35.559821, the results indicate a commendable level of fit.

\section{Conclusion \& Future Work}
In this chapter, we will conclude with the properties of Kalman filtering and gain inspiration through extensive exploration. 

Firstly, our discussion will pivot towards the concept of sequential processing in Kalman filtering. The characteristic of this method is to adopt an incremental data processing strategy, processing one observation at a time. This progressive technology not only has high computational efficiency but also greatly benefits in real-time application and processing of large datasets. Therefore, Kalman filtering has shown commendable performance in fitting SDE systems. Evident from the RMSE values observed in various models, such as the OU process, the OU process with jumps, the Heston model, and the Heston model with jumps, Kalman filtering shows impressive fitting results. Specifically, the RMSE values are 4.560e-05 and 0.0001021 in the OU process, and 1.268415 and 6.37269 in the Heston models, respectively. Intriguingly, Kalman filtering maintains its robust performance even in the presence of jumps in Lévy processes. This characteristic of Kalman filtering offers valuable insights for future research, particularly in addressing noise issues in Lévy spaces.

In addition, we consider closed-form estimation in linear systems. Here, the Kalman filter shines brightly, providing a direct closed-form solution for state estimation. This direct method greatly reduces computational complexity, especially compared to resource-intensive iterative optimization methods. In addition, in terms of computational load, Kalman filters operating in linear systems typically have a lighter burden and are linearly proportional to the amount of data. This is very evident in comparison to MLE. The results of parameter estimation using Kalman filtering and MLE estimation are similar for the same SDE. But the time required for the two methods is completely different. Kalman filter takes 0.30717 seconds and 0.265414 seconds respectively, but MLE is fertilized for 4.418 seconds and 27.83828 seconds. This may be because MLE typically requires more computationally intensive iterative optimization techniques, especially in complex or large-scale models. In summary, the Kalman filtering method is faster in parameter estimation of SDE systems, mainly due to its sequential updating mechanism and the existence of closed-form solutions in some cases. At the same time, MLE requires iterations and may require larger computational complexity.

After that, we discussed the inherent assumptions and constraints. The assumptions of the standard Kalman filter are linear and Gaussian noise. To adapt to nonlinear models, alternative solutions such as EKF or Particle-EKF have begun to take effect. Although these variants may increase computational complexity, they generally maintain their effectiveness and applicability. Meanwhile, we evaluated the robustness of the Kalman filter in SDE systems, and conducted tests on the EKF within the Heston model using various parameters:
\begin{table}[ht]
    \centering
    \begin{tabular}{|c|c|c|c|c|c|c|}
        \hline
        Parameter & $\mu$ & $\kappa$ & $\theta$ & $\sigma$ & $\rho$ & RMSE  \\
        \hline 
        task 1 & 0.04 & 0.3 & 1.5 & -0.6 & 0.04 & 1.268415 \\
        \hline 
        task 2 & 0.1 & 1.0 & 0.02 & 0.1 & -0.8 &  1.01548 \\
        \hline
        task 3 & 0.3 & 2.0 & 0.01 & 0.6 & -0.1 & 0.77223 \\
        \hline
        task 4 & 0.23 & 0.01 & 3 & 0.2 & 0 & 1.66183 \\
        \hline
    \end{tabular}
    \vspace{0.25em}
    \caption{EKF In Heston with Different Parameter}
    \label{tab:ROB}
\end{table}

Based on the above $\bf{Table}$ $\bf{\ref{tab:ROB}}$, We find that all the values of RMSE do not vary outrageously, and almost all of them vary around 1. This suggests that our EKF is stable in the Heston model.

In the next stage, our attention will shift towards a more complex framework, using Chen's model\cite{CHU2009165} as an example. This model is known for its subtle description of interest rate dynamics, providing a three-factor model rather than the two-factor structure of the Heston model we have explored so far. Based on this progress, we plan to rigorously evaluate the adaptability of Kalman filters by integrating jump components, similar to the methods used in the Bates model. This will enable us to more accurately evaluate the effectiveness of Kalman filters in fitting complex financial models.

After this, our research will focus more on prediction. Our main goal is to capitalize on the potential of sliding window techniques as a tool for prediction. We aim to develop a more nuanced understanding of predictive dynamics. This forecasting journey will allow us to rigorously apply the Kalman Filter to the S\&P 500 in particular. Our intention is not just to apply this technique in a theoretical vacuum, but to utilize it in the challenging and unpredictable realm of real-world financial data. This will allow us to carefully evaluate the robustness and predictive power of the filter in an environment filled with complexities. As we progress, our research will extend to evaluating the adaptability and accuracy of Kalman filters under a variety of market conditions, including periods of high volatility and market corrections. We will investigate how the filter responds to sudden economic events and changes in market trends. In addition, we aim to compare the performance of Kalman filters with other forecasting models and assess their relative strengths and weaknesses in predicting market volatility. 

\section*{Github Link}
All code pertaining to this project can be accessed at:

\url{https://github.com/bao021600/Application-of-Kalman-Filter-in-Stochastic-Differentional-Equations}

\bibliographystyle{unsrt}  
\bibliography{references}

\end{document}